\documentclass[11pt,a4paper,english]{article}
\usepackage[T1]{fontenc}
\usepackage[utf8]{inputenc}
\usepackage{babel}
\usepackage{authblk}
\usepackage{moreverb}
\usepackage{graphicx}
\usepackage{float}
\usepackage{balance}
\usepackage[all]{nowidow}
\usepackage{multirow}
\usepackage{xcolor,colortbl}
\usepackage{subcaption}
\usepackage{array}
\usepackage{amssymb}
\usepackage{diagbox}
\usepackage{cite}

\newcommand{\twobar}{//}

\begin{document}

\title{From Zero to Fog: Efficient Engineering of Fog-Based Internet of Things Applications\thanks{Funded by the Deutsche Forschungsgemeinschaft (DFG, German Research Foundation) -- 415899119.\\This work has been published in Wiley -- Software: Practice and Experience vol. 51 no. 8 pp. 1798--1821

    }}

\author{Tobias Pfandzelter}
\author{Jonathan Hasenburg}
\author{David Bermbach}
\affil{Mobile Cloud Computing Research Group\\
    TU Berlin \& ECDF\\
    Berlin, Germany\\
    \texttt{\{tp, jh, db\}@mcc.tu-berlin.de}}

\date{}

\maketitle

\abstract{In IoT data processing, cloud computing alone does not suffice due to latency constraints, bandwidth limitations, and privacy concerns.
    By introducing intermediary nodes closer to the edge of the network that offer compute services in proximity to IoT devices, fog computing can reduce network strain and high access latency to application services.
    While this is the only viable approach to enable efficient IoT applications, the issue of component placement among cloud and intermediary nodes in the fog adds a new dimension to system design.
    State-of-the-art solutions to this issue rely on simulation or solving a formalized assignment problem through heuristics only, which both have their drawbacks.
    In this paper, we present a five-step process for designing practical fog-based IoT applications that combines best practices, simulation, and testbed analysis to converge towards an efficient system architecture.
    We then apply this process in a smart factory case study.
    By deploying filtered options to a physical testbed, we show that each step of our process converges towards more efficient application designs.}

\section{Introduction}
\label{sec:introduction}

For more than a decade, cloud computing has been the dominant paradigm in designing and deploying software services.
However, it is not a good fit for new application domains such as the Internet of Things (IoT).
Sending the world's IoT data to a centralized cloud for processing is both inefficient and prohibitively expensive~\cite{Zhang2015-cb}.
Processing should instead happen where IoT data is generated and needed~\cite{Bermbach2018-bb}.
Fog computing, as first proposed by Bonomi et al.~\cite{Bonomi2012-if}, supplies the necessary paradigm shift:
It extends the cloud to the edge of the network so that applications can leverage additional infrastructure between the cloud and end-devices.
From powerful data centers in larger cities to small, single-board computers co-located with cellular base stations, application designers can deploy their services not only in a central cloud but anywhere on the edge-cloud continuum.
While cloud resources still provide elastic, seemingly infinite scalability at low cost, edge infrastructure offers service consumers low latency access while also consuming less network bandwidth~\cite{Bermbach2018-bb}.
Overall, fog computing enables previously impossible application architectures but at the same time makes application design more complex.
When designing fog-based IoT applications, the placement of software services within the fog is now a new dimension to be considered in addition to actual building of the application.

Correctly placing services, however, is vital in leveraging fog computing for the IoT as it directly influences both the quality and cost of applications.
At the same time, the number of deployment options increases exponentially with each service or location.
Existing approaches to designing fog-based IoT data processing applications all have their drawbacks:
First, there are those that try to parametrize the entire system to form an optimization problem solved algorithmically (e.g.,~\cite{Brogi2017-nl,Skarlat2017-ye,Mahmud2018-ns,Hong2016-hv,Skarlat2017-zx}) or via simulation (e.g.,~\cite{Hasenburg2018-fn,Hasenburg2018-nd,Gupta2017-jx}).
While such approaches are highly valuable for providing insights, their accuracy is inherently limited by the assumptions of the model -- in particular, such approaches cannot capture runtime characteristics of actual software components, which are highly dependent on concrete implementation choices.
Information on implementation choices, however, is not be available at design time and their actual impact depends on the target hardware.
Alternatively, a second approach is to follow guidelines, best practices, or reference architectures such as~\cite{paper_pfandzelter_streams_functions,Gusev2019-ch,Karagiannis2020-kx,Santos2020-qx}.
Although useful as a starting point, these generalized target scenarios are not sufficient for a specific use-case.
Third, there are approaches that introduce tooling to create (emulated) fog testbeds (e.g.,~\cite{Hasenburg2019-er,Mayer2017-dt,Coutinho2018-wa}) to deploy, test, and benchmark applications.
While such experiments can provide the most accurate insights into application behavior, the high level of accompanying effort and running costs make them a poor fit for exploring thousands or millions of deployment options.

In this paper, we propose a new process for designing efficient fog-based systems that combines and extends existing approaches, namely following best practices, simulation, and testbed emulation.
This combination enables us to leverage the advantages of each approach while mitigating their respective limitations.
For instance, we apply best practices to reduce the parameter space for simulation, which prevents incurrence of costs for simulating the entire parameter space without sacrificing the accuracy of simulation results.
Our overall goal is thus to identify an efficient fog application design as effectively as possible.

To this end, we make two core contributions:

\begin{itemize}
    \item  We extend and integrate previous research of ours into a novel framework. We use best practices~\cite{paper_pfandzelter_streams_functions}, simulation with \emph{FogExplorer}~\cite{Hasenburg2018-fn,Hasenburg2018-nd}, and infrastructure emulation with \emph{MockFog}~\cite{Hasenburg2019-er} (Section~\ref{sec:contribution}).
    \item We implement a smart factory application following our proposed process and compare the final application design to a range of discarded design options in experiments on a physical fog testbed (Section~\ref{sec:evaluation}).
\end{itemize}

\section{IoT Applications in the Fog}
\label{sec:background}

In this section, we summarize fog computing concepts and discuss characteristics of fog-based IoT applications and efficient IoT application design.

\subsection{Characteristics and Challenges of Fog Computing}

\begin{figure}
    \centering
    \includegraphics[width=\textwidth]{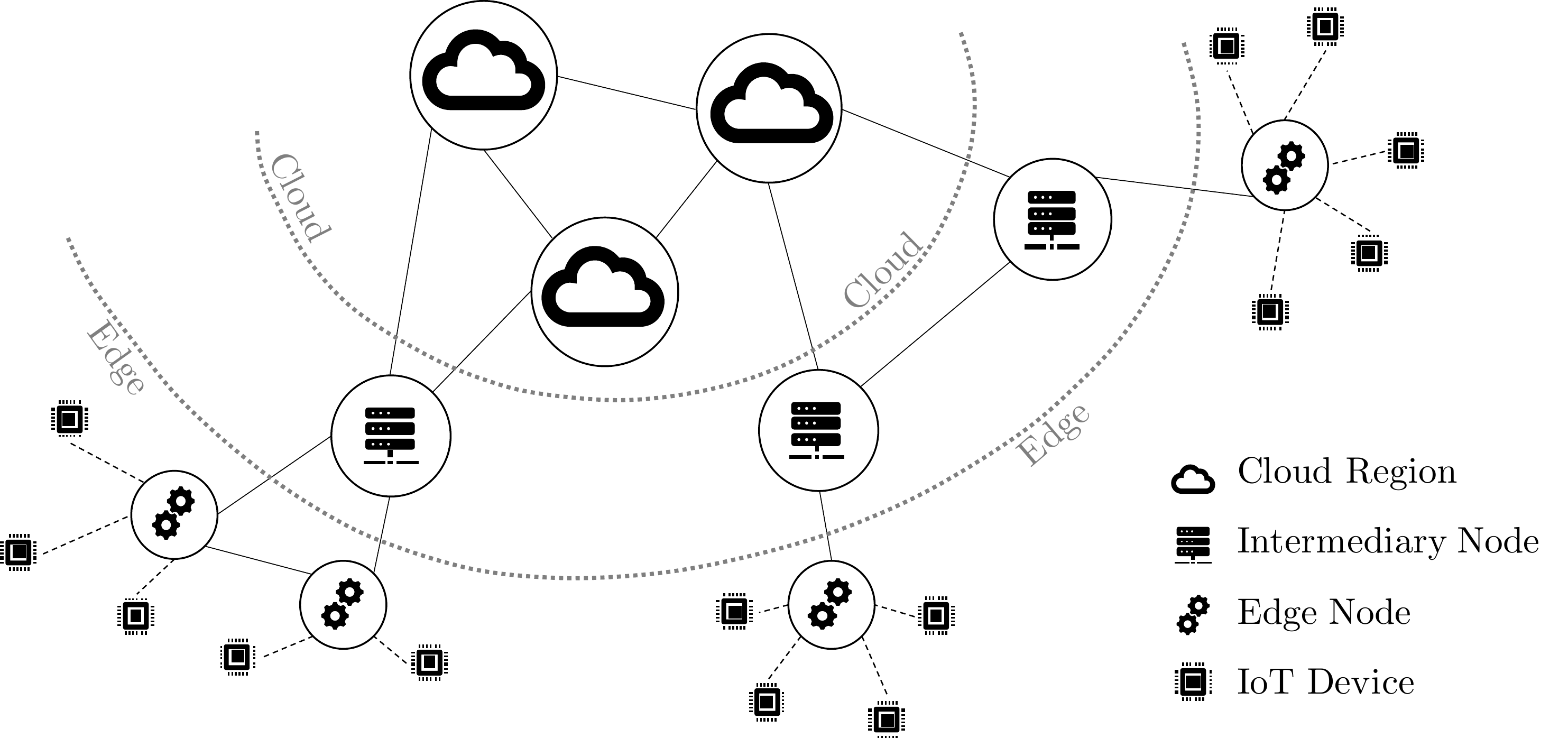}
    \caption{The layered fog architecture comprises cloud, intermediary, and edge nodes, as well as IoT devices.}
    \label{fig:fogarch}
\end{figure}

Our definition of fog computing is adapted from~\cite{Bermbach2018-bb}.
Fog computing is the extension of the cloud toward the edge of the network.
The idea is to distribute applications across a wide variety of infrastructure including cloud resources, intermediary nodes, edge computing, and even on-device computation.
In this way, application developers can leverage both low access latency at the edge and scalability in the cloud.
We show an example of a layered fog architecture in Fig.~\ref{fig:fogarch}.

Because fog computing combines platforms from different vendors (e.g., a cloud provider or a network provider), heterogeneity is a major challenge.
Different platforms are likely to provide different programming models and service levels.
Furthermore, intermediary and especially edge nodes are also likely to be more expensive and less scalable than their cloud counterparts.

A major obstacle to using fog computing is that applications need to be deployed in a distributed manner, with different software components placed on different nodes in the fog.
This is impossible when dealing with traditional monolithic information systems.
Only a modularized application split into distinct software services allows each service to be placed at specific locations within the fog, whether that be in the cloud or toward the edge.
While increasing the communication overhead, smaller services are necessary for fine-granular scaling and enable more flexibility in service placement on the fog infrastructure~\cite{Bermbach2018-bb}.
To this end, leveraging lightweight virtualization technologies such as Docker can make software deployment easier~\cite{Morabito2018-ip}.

\subsection{Data Processing Paradigms in IoT Applications}

IoT applications analyze data from sensors or process them to trigger actor devices and software systems~\cite{paper_pfandzelter_streams_functions}.
A key characteristic of IoT applications is that they do not follow a request-response model as do user-facing systems.
Instead, data move through a processing pipeline in a more ``linear'' way -- typically in the form of a directed acyclic graph (DAG).
Overall, there are two classes of IoT data processing: \emph{event processing} and \emph{data analytics}.
Zhang et al.~\cite{Zhang2015-cb} describe these as ``real-time applications with low-latency requirements'' and ``ambient data collection and analytics,'' respectively.
An application often comprises multiple data processing components that can each be classified individually in this manner.

In event processing, events from the outside world (measured through connected devices) trigger actions in the system and, by extension, possibly in the physical world.
The main focus here is time sensitivity to satisfy tight latency requirements.
Advantageously, operations are thus also well-defined and simple, and events as data points are small as they carry only metadata~\cite{Govindarajan2014-dm}.

Data analytics is the process of collecting and processing data to obtain information.
Complex operations are applied to data from multiple sources over a longer period of time here~\cite{Anawar2018-lv}.

\subsection{Service Level and Cost in Fog Applications}
\label{sec:efficiency}

We consider two dimensions of efficiency in fog-based IoT applications: service level and cost.

\emph{Service level}, often also referred to as \emph{quality of service} (QoS), can be both the availability of the application and the access latency for specific services~\cite{Bermbach2018-bb,benchmarkingbook}.
Access latency is highly dependent on service placement and is determined by data processing and transmission.
Data processing latency describes the time that passes between the input into the processing unit (e.g., a cloud function) and the output of a computed result.
Data transmission latency, on the other hand, is the delay from the first packet of data to be sent by the sender to the last packet of data to be received by the receiver.
Availability of individual services is dependent on two main factors:
software availability and platform availability.
We abstract from software availability in this paper, as software testing is an orthogonal problem to the service placement problem we address.
Platform availability depends on the availability of the network, server, or storage.
Although cloud platforms also suffer from outages periodically, we consider them to have higher availability than edge or fog devices.
Redundancy is managed by the cloud providers and providers are liable for any outages that violate their service level agreements (SLA).

\emph{Cost} is incurred through the usage of resources in the fog such as compute, storage, and network bandwidth and through upfront investment in IoT devices or other hardware.
Generally, compute and storage are far cheaper closer toward the cloud, as providers can leverage economies of scale in large public cloud data centers rather than on the edge, where only privately used devices may be available~\cite{7469991,Bermbach2018-bb}.
For network bandwidth, fog platform providers often charge for outgoing and incoming traffic to a data center and IoT devices may use cellular network access where each packet incurs a specific cost.
These costs are the main contributors towards the total cost of operating an IoT application.

When designing fog-based IoT applications, different design options result in different service levels and cost.
An efficient design offers the best possible QoS levels at the lowest possible cost (i.e., it finds a sweet spot in the QoS and cost tradeoffs~\cite{benchmarkingbook,diss_bermbach,kossmann2010evaluation}, as QoS and cost are not independent of each other).
Deploying powerful servers at every edge location minimizes latency but results in higher costs and lower availability.
Similarly, moving all services to the cloud minimizes cost and maximizes availability but increases access transmission latency at the same time~\cite{Bermbach2018-bb,Zhang2015-cb}.

\section{Designing Efficient IoT Applications}
\label{sec:contribution}

In this section, we present our proposed fog application design process.
We start by giving a high-level overview of our approach before describing the individual steps in detail.

\subsection{Overview of Our Five-Step Process}
\begin{figure}
    \centering
    \includegraphics[width=\textwidth]{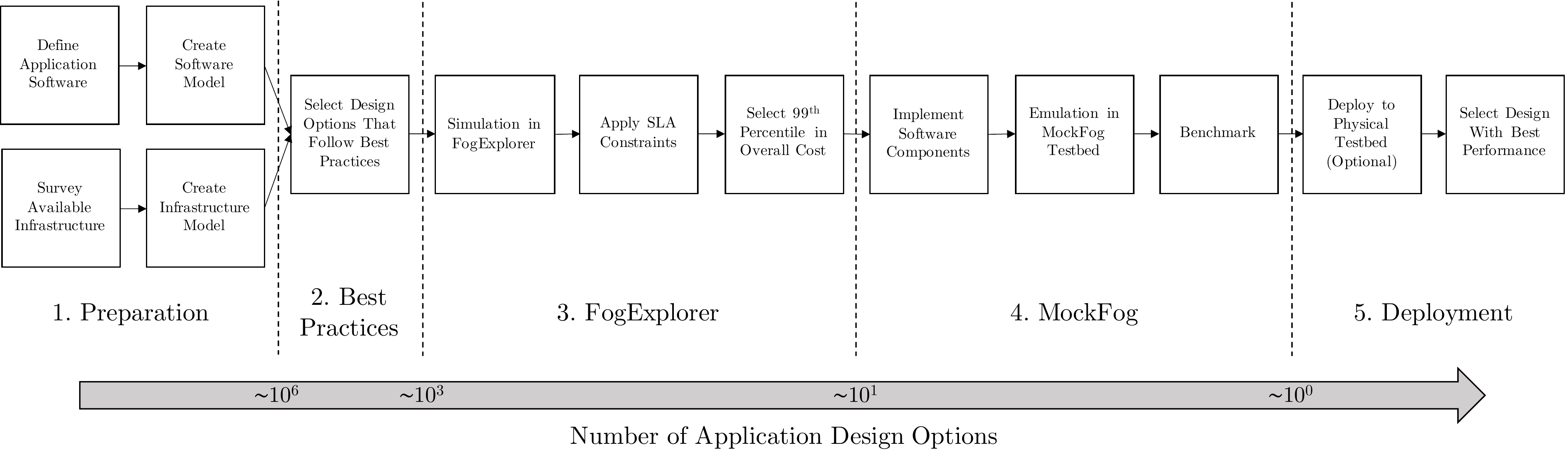}
    \caption{Starting from the set of all possible application design options derived from a software and infrastructure model, we remove poor design options in each step of the application design process to converge on the most efficient solution.}
    \label{fig:process}
\end{figure}
The process we propose for designing efficient fog-based IoT applications comprises five main steps.
Initially, there is a broad range of design options, in the order of $10^{6}$, each of which describes a mapping of software services to nodes in the cloud, edge, or in-between (i.e., the service placement).
Each step of the proposed process then filters out application design options starting with the Cartesian product of all software and infrastructure models, thereby converging on a limited set of most efficient designs.
The key idea is to create a sequence of steps in which each step provides more accurate recommendations than its predecessor but is also more expensive to execute.
Since each step reduces the application design space by orders of magnitude, we use more expensive analysis steps for only a limited number of options late in the process while relying on low-cost heuristics in the first steps (see Figure~\ref{fig:process} for a high-level overview of the proposed process).

In the first step, we build models of software components and the infrastructure on which the application will be deployed.
In each later step, we then extend these models and augment them with additional details as available further in the design process.
Finally, we are able to select an efficient fog application design.

In the second step, we apply a set of best practices in IoT data processing.
By following these informed rules, we can discard all highly inefficient options at this stage.
This reduces the set of options that we have to consider later in the process, enabling us to move through these subsequent steps more efficiently.
As the number of available options grows exponentially with each additional component, this step reduces the design options considered in the subsequent steps from millions ($\sim10^{6}$) to just thousands ($\sim10^{3}$).

In the third step, we simulate service placement to infrastructure components.
This enables us to calculate service cost based on the given cost factors and examine latency constraints for different designs.
By introducing service level objectives (SLOs) for parts of the application, we can remove application design options that violate required service levels and instead focus only on inexpensive options that conform to all constraints, reducing the set of viable application design options to an order of $10^{1}$.

In the fourth step, we set up emulated testbeds for each of the remaining application design options to deploy and benchmark software services.
As this step is expensive and time-consuming, we propose to use only the options in the 95\textsuperscript{th} percentile of the second step, again reducing the number of considered application design options by orders of magnitude.
Based on the number of remaining design options, this selection may be limited or broadened, reducing testing costs or leading to more accurate results, respectively.

This process eventually converges toward a small set of highly efficient design options.
If available, the options that show the best performance at good cost levels can then be deployed on a physical testbed or the actual infrastructure to measure their performance in their real environment (fifth step).

\subsection{Step 1: Software and Infrastructure Models}
\label{sec:models}

Our process requires basic insights into the available runtime infrastructure and the individual software services.
These insights can be provided by domain experts.
For example, system administrators may be able to provide information on available infrastructure, while application developers can identify and classify application services.
We start with a notion of infrastructure components, yet at this early step in the design process we cannot assume that detailed information about runtime infrastructure is available.
We therefore need only high-level, abstract descriptions of available data processing locations (such as IoT devices, edge nodes, or cloud platform providers).
Such knowledge can, for example, be gained by surveying and analyzing eligible providers and products or by comparing options for IoT devices and gateways.
For some more complex use cases, synthesizing possible edge infrastructure configurations as proposed by Rausch et al.~\cite{Rausch2020-cm} could be an alternative approach.
Furthermore, possible infrastructure components should be selected with their availability in mind.
For applications that require high availability, it is helpful to consider only infrastructure components that can provide sufficiently high platform availability.

\begin{figure}
    \centering
    \includegraphics[width=\textwidth]{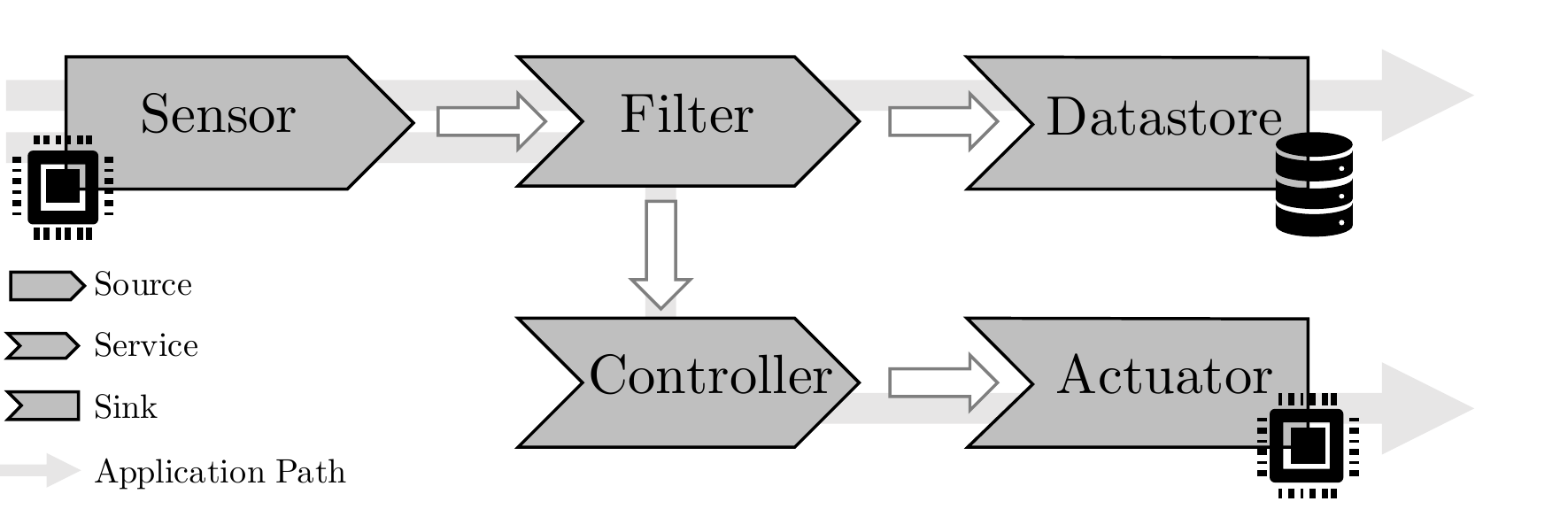}
    \caption{Example software model with two application paths, sources, services, and sinks.}
    \label{fig:applicationcomponents}
\end{figure}

Aside from infrastructure components, we also model software components\footnote{Please, note that we are not trying to derive a formal model as used in either mathematically formulated optimization problems or in standardized modeling languages. Rather, our efforts focus on deriving and abstracting certain properties from an application and its available infrastructure.
    The manner in which this abstract information is represented is irrelevant for our purposes.}
At this point, no actual implementation has to be available yet.
For our model, we use three kinds of components: \emph{sources}, \emph{services}, and \emph{sinks}.
Sources are components that produce new data.
For an IoT use case, sources are typically IoT sensors.
Services consume data and perform operations, thereby producing new data.
Services could, for instance, transform data through aggregation or trigger events.
Finally, sinks are components that persist dat (e.g., a database system), or interact with the physical world based on data (e.g., an IoT actuator).
Sinks that persist data can also have a secondary role as sources exposing historical data.
We show an example application of this kind in Fig.~\ref{fig:applicationcomponents}.
We define the overall application as a collection of \emph{application paths}.
Each application path starts with one or more data sources, has a number of services along the way and ends in one sink (i.e., an application path is the DAG of processing steps that leads to a particular sink).
Again, these application paths can easily be identified by the application developers as they reflect the application's business logic.
At this point, albeit early in the process, it is already useful to simplify both software and infrastructure models.
In most IoT applications, specific components are instances of the same class of components.
In a smart home use case, for example, there could be various light bulbs and corresponding light switches.
Assuming that each switch controls a number of lights, a pattern emerges.
To simplify simulation and benchmarking, we model only one application path and later apply this to all instances of light switches and lights in the system.
This also allows our process to scale well and to require less upfront information about the system, while not influencing the results as we merge instances of the same component rather than modifying them.

For sources and sinks, the mapping to infrastructure components is clear, as these are tied to the physical world.
An IoT device, for instance, exists as a physical device (i.e., an infrastructure component) and as a source in the software model.
Consequently, we only need to consider the placement of services (i.e., the software components that process data) in the subsequent steps of the design process.

\subsection{Step 2: Applying Best Practices}
\label{sec:bestpractices}

In previous work~\cite{paper_pfandzelter_streams_functions}, we proposed best practices for fog-based IoT application design, which we now use to exclude unsuitable application design options.
In the following, we will briefly describe how we apply these best practices, which we split into rules for event processing and data analytics application paths.

In event processing, processing is time-sensitive and services should be placed on the shortest communication path between data source(s) and sink, as close to the cloud as possible to minimize cost, and as close to the edge as necessary to fulfill SLOs.
As processing a single event is not compute-intensive, minimizing round-trip time is more important than reducing processing delay, especially for reactive and real-time systems~\cite{Manna2012-cj}.
However, as cloud computing resources scale better and moving toward the edge reduces flexibility and increases cost, it is still important to process events as close to the cloud as possible.
That means selecting the infrastructure node that provides the most flexibility and least expensive compute power from the set of nodes on the shortest path between the event source and its sink.

In data analytics, however, time sensitivity is not the main priority.
Operations are complex and require a lot of processing power.
These operations range from filtering or aggregation to predictive analytics with machine learning.
Furthermore, services here must consider and even combine data from different sources.
This also includes complex event processing, where events from thousands of different sources, such as IoT sensors, are analyzed in an aggregated manner.
On the cloud-edge continuum, data analytics processors that preprocess data should be kept as close to the edge as possible to reduce data volume on the network, yet also as close to the cloud as necessary given their computational complexity.
Compute-heavy operators, on the other hand, should be placed near the cloud, where processing is cheaper.

Given these best practices, we can filter the set of application design options.
When filtering, we consider each option individually.
First, we identify whether each application path targets event processing or data analytics.
For an event processing application path, infrastructure nodes located on the shortest path between the infrastructure components that host the event source and sink are an efficient location for software services.
In data processing, we favor preprocessing of data close to the edge where possible.
This reduces usage of bandwidth toward the cloud, where we propose to place more complex data processing.
We also rule out options where the resulting data flow uses the same network links more than twice.

\subsection{Step 3: Simulation}
\label{sec:fogexplorer}

In the third step, we use simulation to analyze the remaining application design options.
The simulation is conducted using FogExplorer~\cite{Hasenburg2018-fn,Hasenburg2018-nd}, which we presented in previous work.
FogExplorer can be used in an interactive way, with application designers able to update mappings and observe the resulting metric values instantly.
Alternatively, FogExplorer can also be used in a batch mode through its API.

Based on an infrastructure and software model, FogExplorer calculates four metrics per mapping: \emph{processing cost}, \emph{processing time}, \emph{transmission cost}, and \emph{transmission time}.
Processing cost and transmission cost describe the average cost per second within the system.
Processing time and transmission time describe latency induced by services and transmission of data.

To calculate these metrics, FogExplorer first determines the data stream routing by identifying the path with the lowest total bandwidth cost for each set of two communicating software components.
In a second step, FogExplorer calculates resource usage to assert that the selected mapping does not exceed resource limits.
For example, a connection may have a limited amount of bandwidth.
In this case, FogExplorer will determine if the bandwidth required by any connection within the mapping exceeds the available bandwidth.
In the third step, FogExplorer calculates total cost based on resource usage.
Transmission costs depend on bandwidth used and the respective bandwidth price.
In a similar manner, FogExplorer calculates processing costs.
In addition, FogExplorer also determines time metrics and calculates processing time and transmission time for each application path.
Processing time is the total latency induced by services processing data, while transmission time is the total connection latency along the application path.
Finally, FogExplorer tallies transmission costs and processing costs, as well as transmission times and processing times to project the total cost and end-to-end latency of the given mapping.

We use FogExplorer to further filter out application design options as the third step of our proposed process.
To use this simulation, we have to extend our software and infrastructure models slightly.

In the infrastructure model, we also specify different hardware options that are available for each node.
At an edge data center location, for instance, it may be possible to install different types of servers with different capabilities as well as different price points.
Here, FogExplorer allows us to compare these different options to find the most efficient one.
Although this increases the space of application design options, it is necessary to determine the optimal infrastructure.
For each infrastructure option at each node, we specify a \emph{relativePerformanceIndicator}, which is a rough estimate of compute power compared to a chosen reference machine.
For instance, if a machine type has a performance indicator of 2, it is twice as ``fast'' as the reference machine.
Furthermore, the \emph{availableMemory} metric specifies how much memory is available for the machine and the \emph{price} metric specifies the price for using the machine.
Network components are extended with an \emph{availableBandwidth}, a \emph{bandwidthPrice}, and a \emph{latency} for each connection.
If latency cannot be accurately benchmarked ahead of time, it is also possible to use estimates based on link layer performance and geographical locations of nodes as done in~\cite{Hasenburg2020-xi}, for example.

\begin{figure}
    \centering
    \includegraphics[width=\textwidth]{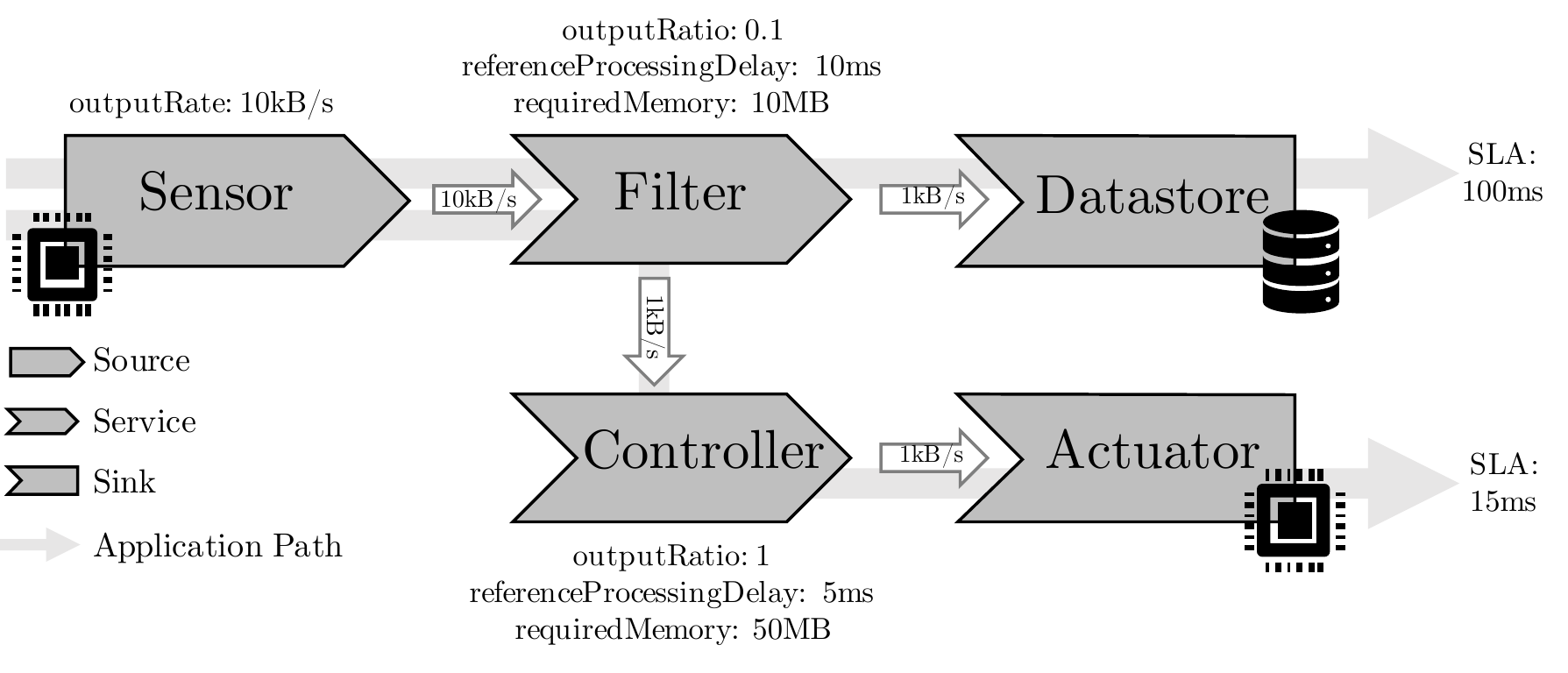}
    \caption{Extension of the example software model; we infer connection data rates from given \emph{outputRates} and \emph{outputRatios}.}
    \label{fig:simulation-applicationcomponents}
\end{figure}

Similarly, we add quantitative attributes to software model components as well.
Sources produce data at a constant rate that we mark as their average \emph{outputRate} in the form of byte/s.
The rate at which services output data depends on their input rate, hence we use an \emph{outputRatio} to calculate their outgoing bandwidth.
For services, we also employ a \emph{referenceProcessingDelay} factor that describes how long, on average, the service needs to process data on the aforementioned reference machine, and a \emph{requiredMemory} metric to describe the amount of memory needed by the service.
Of course, both sinks and sources as software components require a certain amount of memory as well once they are running on an infrastructure node.
The infrastructure nodes then incur costs for running these components.
As we have described, however, mapping for sources and sinks is fixed, as these components relate to objects in the physical world.
Accordingly, while it is possible to simulate costs incurred here as well, these costs would be static and, subsequently, not influence our decision on one application design option versus another.
This is why we omit them in the simulation and focus exclusively on resources required by service components.
The extended version of our example software model from Section~\ref{sec:contribution} is depicted in Fig.~\ref{fig:simulation-applicationcomponents}.

We also introduce SLOs in the form of limits to end-to-end latency for each application path at this point.
As we have described in Section \ref{sec:efficiency}, we measure efficiency for fog application design in cost and latency.
Yet because cost and latency depend on each other, finding the most efficient application design is a difficult multi-objective optimization problem.
Rather than finding the quantitatively optimal solution, we apply constraints in the form of SLOs to convert this problem into single-objective optimization problem\footnote{An alternative would be to use a utility function to transform the multi-objective optimization problem into a single-object optimization problem.}.
While it depends on the specific application, the economic law of diminishing returns usually also applies to the tradeoff between cost and latency~\cite{Mohan2020-cn}.
For example, imagine both a user-facing web service and a machine-to-machine communication use case.
In the first use case, investing a considerable sum to decrease latency by 10ms would often not be useful, but it could be in the second scenario.
Application designers can set the required access latency for all application paths arbitrarily high or low as is required by the application.
The actual limits depend entirely on the business logic and required safety or performance objectives and our process will optimize cost within these specified service levels.

Given these limits on end-to-end latency, we only further consider those models that satisfy these constraints in an efficient way (i.e., at low cost).
From the set of application design options, we select only those that do not violate the service levels for any application path as defined in Section~\ref{sec:models}.
If no model conforms to these constraints, it is useful to reconsider the constraints or available infrastructure.
From the remaining design options, we now select those that we will consider in the testbed step in light of the remaining influence factor (i.e., total cost).
As testbed evaluation is expensive and time-consuming, the number of application design options that will be benchmarked needs to be low.
On the other hand, the design options that are identified as good options in the simulation step are not necessarily the best options (i.e., it can be beneficial to proceed with a broader variety of options).
We propose to solve this tradeoff by proceeding with design options that lie in the 95\textsuperscript{th} percentile when considering their total cost (i.e., the top 5\% design options that have the lowest total cost).
If necessary, this range can be adapted.

\subsection{Step 4: Experiments on an Emulated Testbed}
\label{sec:mockfog}

In the fourth step of our process, we evaluate design options using experiments on an emulated fog testbed.
This evaluation requires an implementation of the application software that we can deploy to the testbed and is thus the most time consuming and costly.
However, the low number of viable options that remain after the first three steps of our process limits the required experiments.
Furthermore, it also limits the implementation effort required as services only need to be implemented for the platforms they can be deployed on in the remaining application design options~\cite{Skarlat2018-pl,Morabito2018-ip}.

To benchmark fog application design options, we propose using MockFog as we presented in~\cite{Hasenburg2019-er}.
MockFog provides an emulated yet realistic environment for functional testing and benchmarking of fog applications in the cloud.
In MockFog, cloud, edge, and intermediate nodes as well as IoT devices are instantiated as cloud virtual machines.
Compute power, memory, and intra-node network characteristics such as latency or failures rates can be configured.
Failure scenarios can be emulated as well.

Once again, we need to modify our initial software and infrastructure models to fit the model used by MockFog.
Instead of a performance indicator as given for machines in the infrastructure model, we now need to quantify the actual \emph{compute power}, \emph{memory}, and \emph{storage} capabilities.
Furthermore, we have to define bandwidth and latency parameters for network connections.
MockFog introduces routers between connected machines rather than direct connections.
Hence, in order for all nodes to be able to communicate, we have to add these routing components where applicable.

Rather than extending the application model, we need to replace it with actual implementations of service and sink components that we then deploy on the MockFog testbed.
For source components, the majority of which are IoT devices, implementation is more difficult.
These source components need to produce IoT data in conformance with the application model.
It is possible to use traces of real IoT data (e.g., through BenchFoundry~\cite{paper_bermbach_benchfoundry}) or to attach real world IoT devices, although this requires consideration of network conditions between these devices and the MockFog testbed location.
Finally, we can also employ artificial workload generators such as Apache JMeter\footnote{https:\twobar{}jmeter.apache.org} to generate data.

On the emulated MockFog testbed we can then analyze the behavior of the IoT application, especially in the context of component placement.
While the MockFog environment also allows us to change configuration parameters at runtime (e.g., to inject failures), we use it only to benchmark application designs under the assumption that the provided application implementation is correct.

\subsection{Step 5: Selection and Deployment}
\label{sec:resolution}

After these four steps, an informed decision on the best design option can be made.
The selected design option is likely to be the most efficient one regarding cost and service level, as it has been selected through best practices and simulation as well as verified on an emulated fog testbed.
If in doubt, the best two or three options can then also be test-deployed in the real runtime environment or on a physical testbed to further substantiate the results.

\section{Evaluation}
\label{sec:evaluation}
To evaluate our approach, we use a case study based on a smart factory scenario.
In the first part, we follow the process described in Section~\ref{sec:contribution} to show that it can be used in practice.
We make all software we use available as open source\footnote{https:\twobar{}github.com/pfandzelter/zero2fog-artifacts}.

In the second part (Section~\ref{subsec:results}), we show that the design option identified by our process is among the best options.
For this purpose, we implement the design on a physical testbed and compare it to alternative design options.
Due to the number of permutations and the resulting experiment effort, it is not feasible to show that the identified option is \emph{the} best option.
We therefore rely on sampling and run experiments with randomly selected design options that we discarded in earlier process steps.

\subsection{Case Study}
\label{subsec:casestudy}

In our case study, we apply our proposed process to a smart factory IoT application.
We start by describing the scenario and deriving software and infrastructure models (Section~\ref{subsubsec:example}), applying our set of best practices (Section~\ref{subsubsec:bestpractices}), using simulation (Section~\ref{subsubsec:simulation}) and testbed experiments with the implemented software services (Section~\ref{subsubsec:testbed}) to identify good design options, and briefly discussing the results of our approach (Section~\ref{subsubsec:casestudyresults}).
This shows that it is indeed possible to pursue our proposed process and to pick a resulting design option.

\subsubsection{Smart Factory IoT Application}
\label{subsubsec:example}

\begin{figure}
    \centering
    \includegraphics[width=\textwidth]{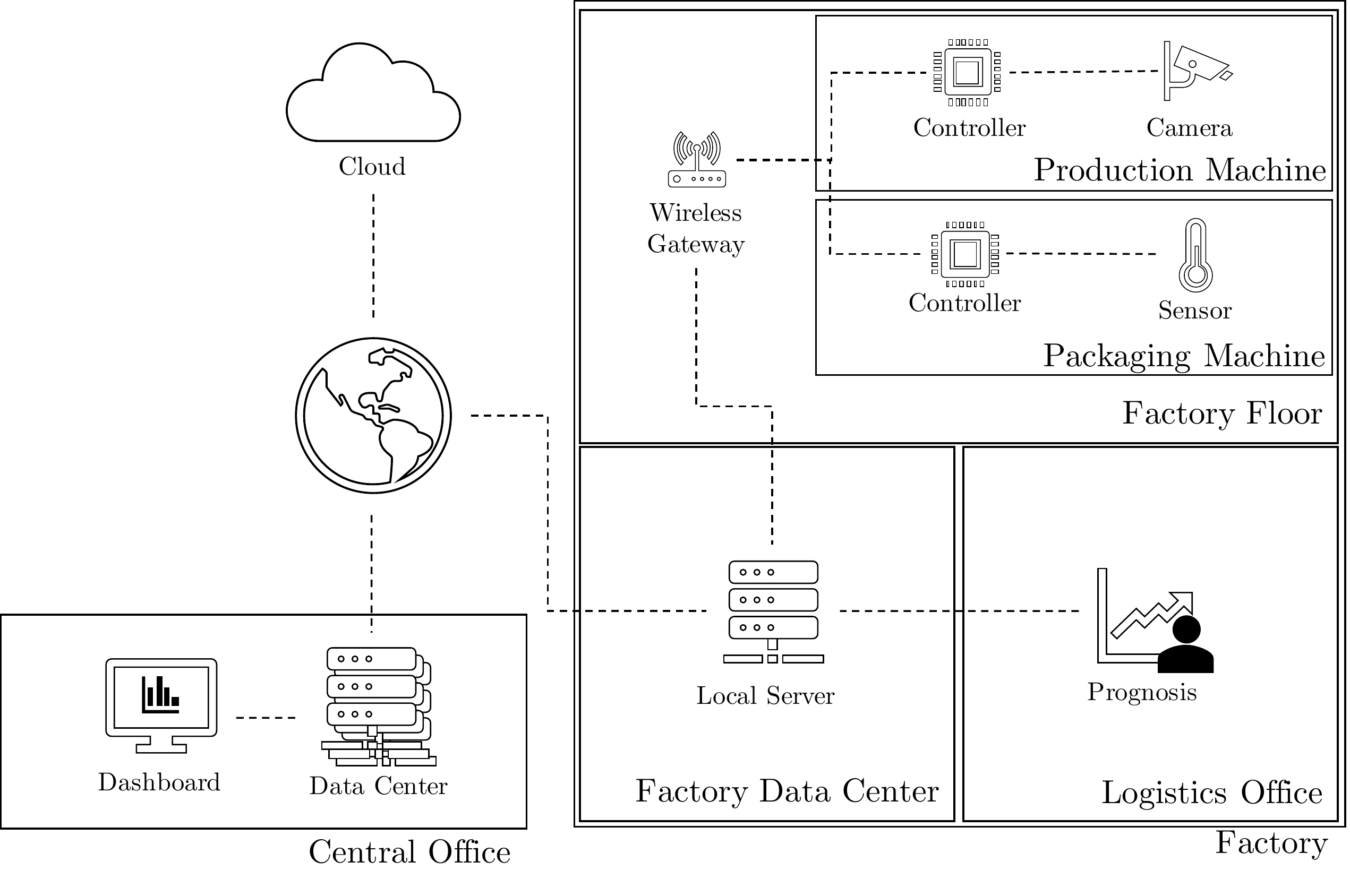}
    \caption{The smart factory comprises a factory floor, factory data center, and logistics office, and is augmented by an office data center and the cloud.}
    \label{fig:overview}
\end{figure}

We provide an overview of our IoT application's components in Figure~\ref{fig:overview}.
The factory comprises a factory floor, a small data center, and a logistics office.
In addition to the factory, there is a central office in an offsite location.

The factory floor has two machines: the \emph{Production Machine} produces a part that the \emph{Packaging Machine} then prepares for shipment.
To ensure that the Packaging Machine processes only faultless parts, the Production Machine has an attached camera that takes a picture of each produced part and checks for defects.
The Packaging Machine should adapt its speed to the output rate of the preceding machine.
Furthermore, the Packaging Machine can only operate within a fixed ambient temperature range and thus has a temperature sensor installed that will shut it off if necessary.
Each machine is also equipped with a controller that controls the speed at which the machine operates.
These controllers are able to communicate over a common wireless gateway.
In the onsite logistics office, logistics personnel decide when to arrange outgoing product shipments.
To this end, a logistics dashboard predicts machine output based on recent productivity.
The factory data center provides some compute power and a connection to the WAN.
In the central company office in an offsite location, the business requires central reporting of factory productivity.
This central office also has a collocated medium-size datacenter.
Additionally, it is possible to leverage cloud computing to outsource some computational tasks.
\begin{figure}
    \centering
    \includegraphics[width=\textwidth]{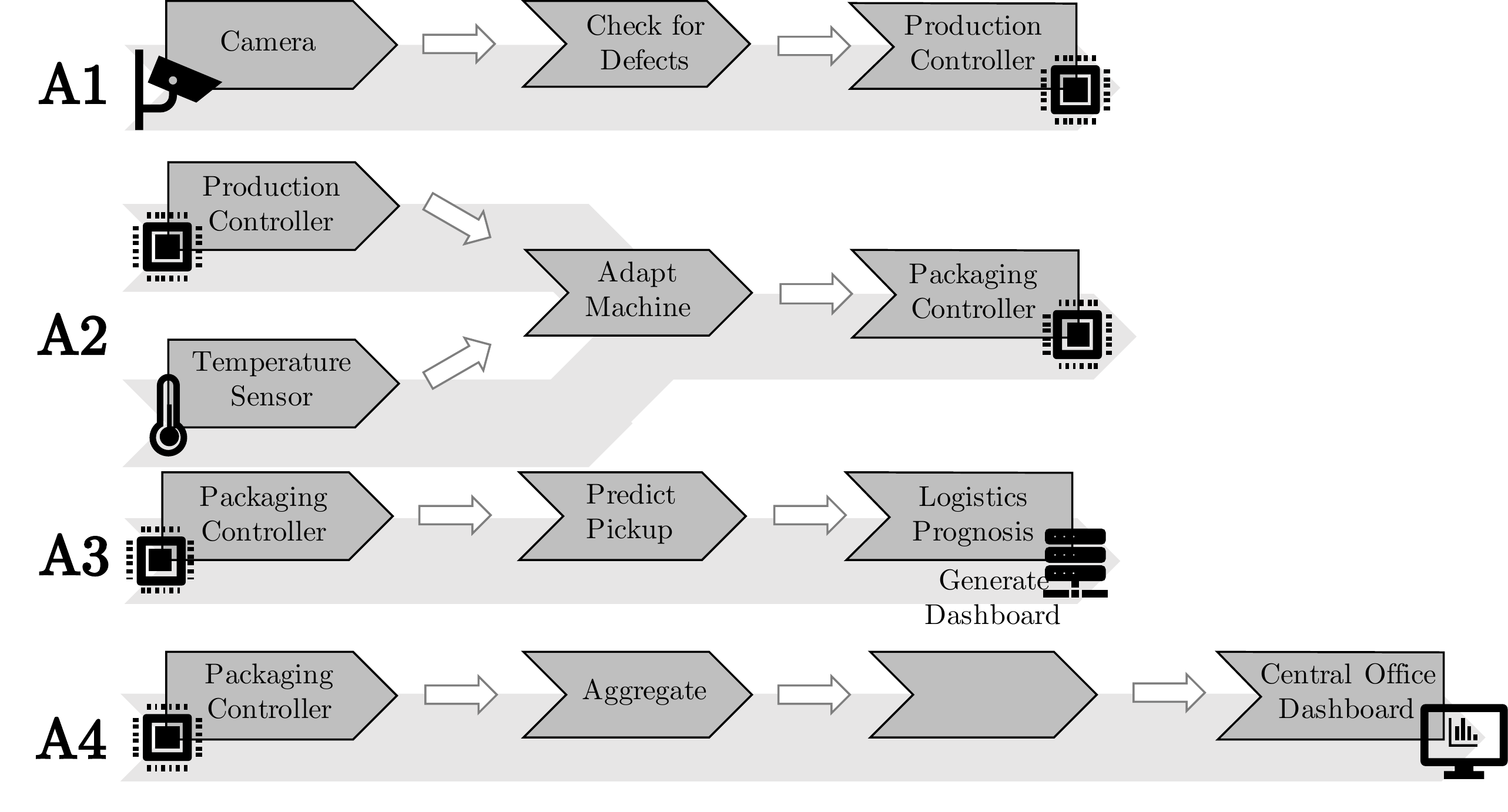}
    \caption{Data sources, services, and sinks in our application. We mark application paths \textbf{A1}-\textbf{A4} for the components.}
    \label{fig:application}
\end{figure}
We use this information to create our infrastructure model with the cloud, data centers in the smart factory and central office as well as wireless gateway, machine controllers, and sensor nodes that all have additional compute capabilities.
We also derive the following application paths from the initial concept (see Figure~\ref{fig:application} for the software model):
\smallskip \\
\noindent
\textbf{A1:} The \emph{Camera} takes pictures of parts leaving the first machine and the \emph{Check for Defects} service analyzes each picture for defects.
In case of a defect, the service instructs the \emph{Production Controller} to discard the respective part.
\smallskip \\
\noindent
\textbf{A2:} The Production Controller has information on the output rate of the machine that produces parts and uses this information to adapt the packaging rate of the packaging machine through an intermediary service.
As a second input, the \emph{Packaging Controller} also relies on data from the \emph{Temperature Sensor} to control the packaging rate. When temperature readings leave a specified range, as detected by the \emph{Adapt Machine} service, the Packaging Controller instructs the packaging machine to pause operation.
\smallskip \\
\noindent
\textbf{A3:} The Packaging Controller provides data on the rate and amount of packaged parts to the \emph{Predict Pickup} service that feeds into the \emph{Logistics Team Prognosis}.
\smallskip \\
\noindent
\textbf{A4:} Data from the Packaging Controller is also consumed by a service that aggregates and filters that data to generate a dashboard for the central office, which then runs inside a browser on a machine in the central office.

Data sources and sinks closely mirror the real world and placement for them is straightforward.
For example, the Camera component in both the infrastructure and software models is the same device as in the real world.
For services, however, we still need to find an efficient mapping.
To this end, we now follow the process introduced in Section~\ref{sec:contribution}.

\subsubsection{Applying Best Practices}
\label{subsubsec:bestpractices}

As described in Section~\ref{sec:bestpractices}, we need to consider all application paths individually in this step.
We begin by classifying each application path and then use the corresponding best practice advice to filter out some application design options.
\smallskip\\
\noindent
\textbf{A1:} Although a photo is larger than a sensor value, we classify A1 as event processing.
Each photo corresponds to an event in the physical world, in this case the production of a part.
The Camera translates this event into a message carrying metadata in the form of an image.
Processing the image is also time-critical as the Production Machine needs to discard any faulty parts before they arrive at the Packaging Machine.
Although the event message has a relatively large size, the Check for Defects service on this application path only needs to consider one source at a time, which, depending on the complexity of analysis for each event, limits processing time.
As such, limited bandwidth and high network latency can be bigger factors in not achieving QoS goals here.
Therefore, image processing should at least be kept on factory premises, or even inside the machine on either the Camera or Production Controller.
A more specific decision is not possible as long as more detailed information about service complexity and infrastructure capabilities is not available at this stage.
\smallskip \\
\noindent
\textbf{A2:} We can make a similar argument for A2.
Here, two event sources produce events independently but a single service that controls the packaging rate consumes all of them.
Again, we classify this path as event processing as events are small in size and decisions need to be made quickly.
Although consuming two data sources, service complexity is also low as  the service does not consider historic data and performs simple calculations.
Thus, placing the Adapt Machine service on factory premises, close to data sources and sinks, is the most efficient option.
\smallskip \\
\noindent
\textbf{A3:} Despite using only one data source producing rather simple data items, we classify A3 as data analytics since it needs to consider current and historical data.
In addition, the processing is more complex as the goal is to predict future packaging rates.
Furthermore, QoS limits for latency are in the range of seconds (rather than milliseconds) as the staff will only periodically check the report.
Consequently, depending on prediction complexity, we propose placing the Predict Pickup service where compute power is the cheapest, in the cloud or a data center for instance.
Correct placement then comes down to a cost calculation between bandwidth price and compute costs, and is part of the subsequent simulation.
\smallskip \\
\noindent
\textbf{A4:} Finally there is A4 which monitors the factory output rate to feed data into a dashboard in the central office.
This is also a data analytics workflow and there are no strict latency constraints.
Instead, data amount and processing complexity are again the limiting factors.
Consequently, as the Aggregate service is a preprocessing step, placing it close to the Packaging Controller limits bandwidth usage.
Similar to A3, we can then place the complex processing service Generate Dashboard where processing is available for the lowest price, which is likely to be the cloud or one of the data centers.

Starting with five services that we can deploy to one of eight infrastructure components each, there are thus 32,768 permutations, and that number grows exponentially with additional services or infrastructure components.
By following our best practices, we managed to reduce the set of options to only 864.

\subsubsection{Application Simulation}
\label{subsubsec:simulation}

\begin{figure}
    \centering
    \includegraphics[width=\textwidth]{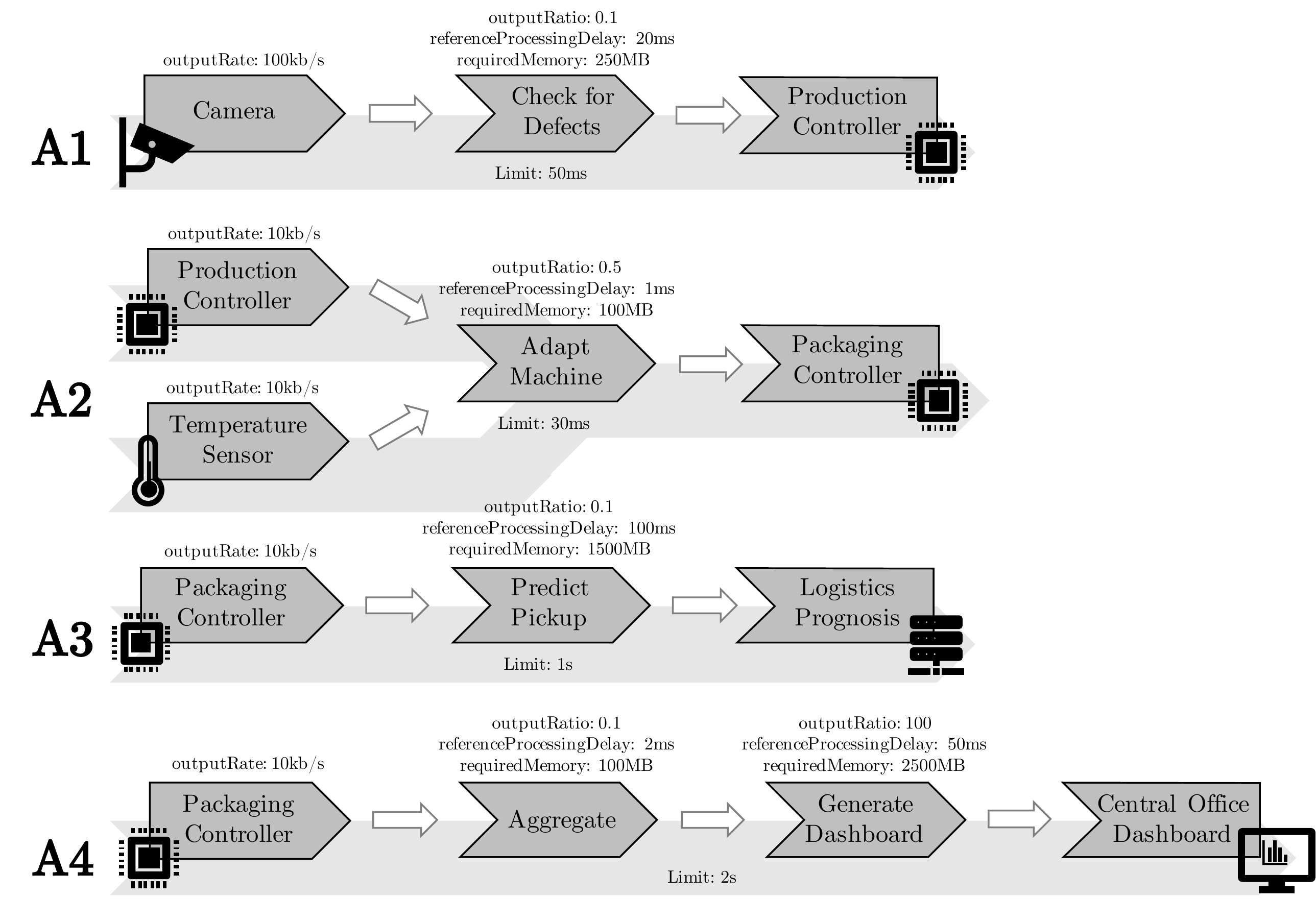}
    \caption{We extend components of the application paths in our software model with attributes as required by FogExplorer: sources have an \emph{outputRate} and services have an \emph{outputRatio}, \emph{referenceProcessingDelay}, and \emph{requiredMemory}. Furthermore, application paths have a QoS limit of acceptable end-to-end latency.}
    \label{fig:fogexplorer_application}
\end{figure}

\begin{figure}
    \centering
    \includegraphics[width=\textwidth]{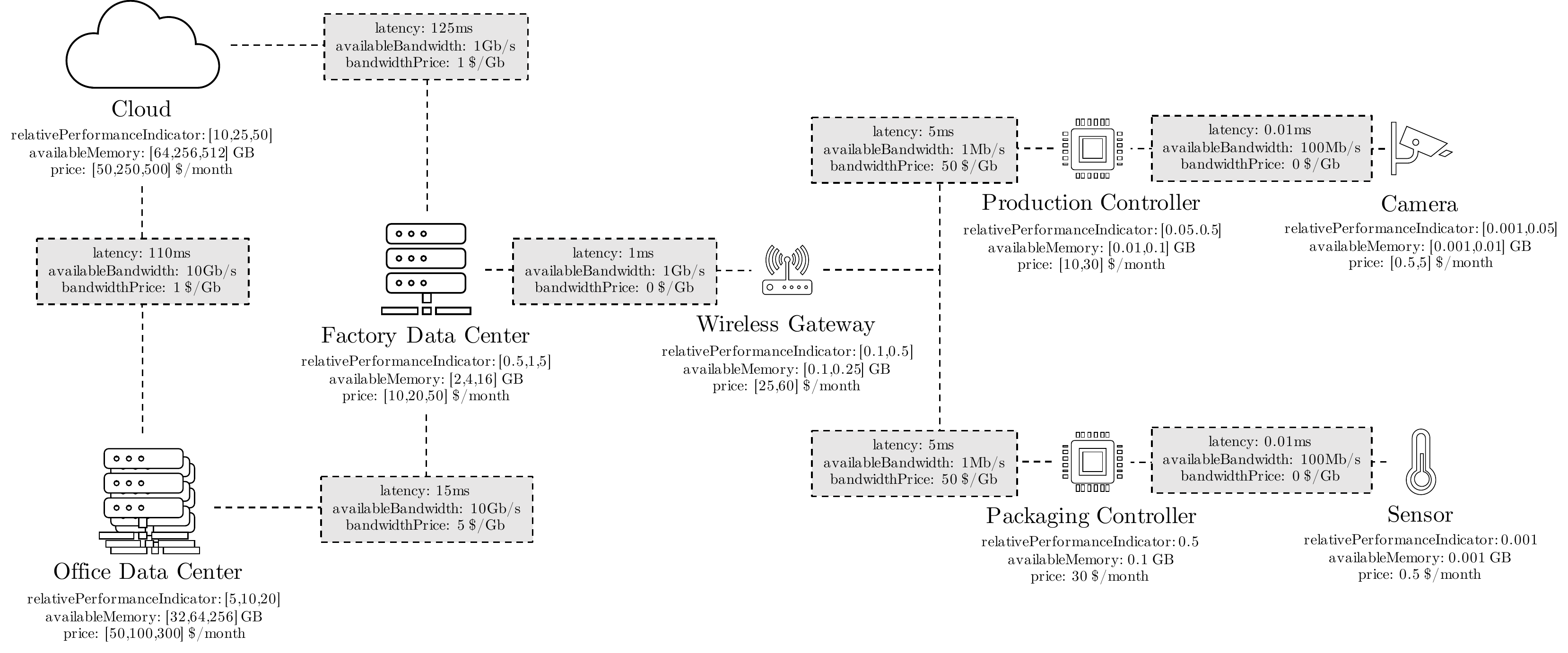}
    \caption{We extend infrastructure components and their network links with more attributes as required by FogExplorer: each node has a \emph{relativePerformanceIndicator}, \emph{availableMemory}, and \emph{MemoryPrice}. Network connections have a \emph{latency}, \emph{availableBandwidth}, and a \emph{bandwidthPrice}. Square brackets denote that more than one hardware option is available at a specific node. These hardware options differ in price and capability.}
    \label{fig:fogexplorer_infrastructure}
\end{figure}

We now use FogExplorer to simulate QoS and the cost of the remaining application design options as explained in Section~\ref{sec:fogexplorer}.
To use FogExplorer, we first need to extend the software and infrastructure models shown in Figures~\ref{fig:fogexplorer_application} and~\ref{fig:fogexplorer_infrastructure}, respectively.
To give an example, in the application model the camera produces data in the form of images at a rate of 100kb/s, and the subsequent service takes an estimated 20ms to process data items on a reference machine with an \emph{outputRatio} of 0.1, meaning that with a 100kb/s input it outputs 10kb/s.
Furthermore, this service requires 250MB of memory.
For each application path we also introduced QoS requirements in the form of latency limits.
In the simulation, we discard any service mapping that violates either of these conditions.
For the A1 application path, for instance, we set an upper limit of 50ms as the delay between taking a picture and the command reaching the production controller.
In the infrastructure model, we introduce different machine options with different capabilities and price points for some nodes.
For example, there are two options for the camera component:
One has computational capabilities of 0.1\% that of the reference machine with 1MB of memory at a price of \$0.5/month while the other has 5\% of the performance of the reference machine with 10MB of memory available at a higher price of \$5/month.

As our case study is fictional, we estimate these prices in lieu of actual infrastructure.
As a basis, we use pricing for a moderate compute instance with a 2-core processor and 4GB of memory on Amazon Web Services (AWS) Lightsail\footnote{https:\twobar{}aws.amazon.com/lightsail}, which costs \$20/month.
This is similar in price and performance to the medium machine option for the Factory Data Center node.
We estimate total cost of ownership per performance to be lower near the cloud and with more powerful machine options, but higher near the edge where maintenance is a greater factor, and extrapolate accordingly.
The A2 application path has two sources and, depending on its placement, these sources have a different connection latency to their common service.
As both sources send their data in parallel, we consider the maximum end-to-end latency for this application path and assert that this does not violate the QoS.

We automate the simulation using the \emph{Node.js} interface of FogExplorer.
Although the number of possible application design options grows exponentially with software and infrastructure components and machine options for nodes, the preceding step in which we discarded options using best practices already limited those options, allowing us to simulate all remaining design options efficiently.
In fact, with current software and infrastructure models we need to consider only 186,624 different options and are able to simulate and calculate metrics for all of them in about one minute on a standard laptop computer.
For comparison purposes, and to emphasize the importance of the first step of our process, there is a total of 7,077,888 application design options and a complete simulation of those takes 50 minutes just for this simple use case.
As such, using only simulation without applying best practices first is infeasible especially for more complex application scenarios.

In addition to overall cost and time metrics, we also calculate metrics for each application path on its own.
This helps us discard options that violate SLO limits.
From 186,624 possible application design options only 2520 are valid and only 215 remain after applying the latency limits we defined.
Consequently, FogExplorer lets us discard the 99.9\% of application design options that are impossible to deploy in practice given infrastructure and SLO constraints.

The options that remain are therefore those that conform to all infrastructure and QoS constraints and we can now choose those that have the lowest overall cost according to the simulation.
We select the application designs in the 95\textsuperscript{th} percentile in the pool of options based on cost, a total of ten designs.
From the simulation, it is clear that placing the Check for Defects service of the A1 application path in the Factory Data Center, the Adapt service of the A2 path on the Packaging Controller or the Factory Data Center, and the Aggregate service of path A4 on the Wireless Gateway are the most efficient application design options.
Furthermore, it becomes apparent that the Camera, Production Controller, and Sensor do not require additional compute capabilities as they do not need to run any data processing services.
For the Factory Data Center, the simulation recommends the medium machine option for each application design options and the least expensive options for both Office Data Center and Cloud.

\subsubsection{Emulated Testbed}
\label{subsubsec:testbed}

Based on the simulation, we chose the ten most efficient application designs and can now deploy these on an emulated MockFog testbed.
Before deployment can begin, we must first implement our software components.
To this end, we implement each source, service, and sink in Go 1.14.
We then install the compiled binaries on the MockFog nodes as Docker containers.
We use an extended version of MockFog for our experiments that is available with all other software artifacts.

\begin{table}[!t]
    \renewcommand{\arraystretch}{1.3}
    \caption{\emph{referencePerformanceIndicator} (rPI) and Corresponding AWS EC2 Instance Types Used in Our MockFog Experiments}
    \label{tab:instance_types}
    \centering
    \begin{tabular}{c|c|r|r|r}
        \textbf{rPI}               & \textbf{\begin{tabular}[c]{@{}c@{}}EC2 Instance\\Type\end{tabular}} & \textbf{\#vCPUs} & \textbf{\begin{tabular}[c]{@{}c@{}}Memory\\in GiB\end{tabular}} & \textbf{\begin{tabular}[c]{@{}c@{}}Median sysbench\\CPU Speed\end{tabular}} \\ \hline
        \(\left[0,1\right[\)       & t2.micro                           & 1                & 2                                  & 1.25                               \\
        \(\left[1,5\right[\)       & t2.medium                          & 2                & 4                                  & 2.90                               \\
        \(\left[5,10\right[\)      & t2.xlarge                          & 4                & 16                                 & 5.89                               \\
        \(\left[10,20\right[\)     & t2.2xlarge                         & 8                & 32                                 & 11.78                              \\
        \(\left[20,50\right[\)     & m5a.12xlarge                       & 48               & 192                                & 45.48                              \\
        \(\left[50,\infty\right[\) & m5a.24xlarge                       & 96               & 384                                & 90.91                              \\
    \end{tabular}
\end{table}

Each node in the system maps to one instance on AWS Elastic Compute Cloud (EC2)\footnote{https:\twobar{}aws.amazon.com/ec2} in the same availability zone of the \emph{us-east-1} region.
To emulate different kinds of hardware, we use different instance types.
We show the mapping from \emph{referencePerformanceIndicator} as employed in FogExplorer to EC2 instance types in Table~\ref{tab:instance_types}.
For instances of the \emph{t2} family, we enable unlimited accrual of CPU credits to prevent inconsistent CPU bursting.
Given the limited number of available instance types, however, this is not as fine-grained as the \emph{referencePerformanceIndicator} in FogExplorer.
It is also not possible to set the \emph{availableMemory} to the same value as in the FogExplorer infrastructure model.
To validate performance differences between instance types we use the \emph{sysbench} CPU benchmark in version 1.0.20.\footnote{https:\twobar{}github.com/akopytov/sysbench/tree/1.0}
This benchmark calculates all prime numbers up to a certain limit, which we set at 1,000,000, in 1,024 threads simultaneously.
It then reports a \emph{CPU speed} metric that describes the number of events the benchmarked CPU was able to handle per second, with each event corresponding to one completed prime computation.
We repeat this benchmark three times and report median results.
As shown in Table~\ref{tab:instance_types}, this metric scales nearly linearly with the amount of CPU cores.
Note that in order to leverage this performance for our application, the services we deploy have to actually use all available CPU cores.
To this end, our implemented application services use multithreading through \emph{goroutines}.
Nevertheless, we can expect that performance does not scale strictly linearly with the number of CPU cores in practice.

\begin{figure}
    \centering
    \begin{subfigure}{0.45\textwidth}
        \includegraphics[width=\linewidth]{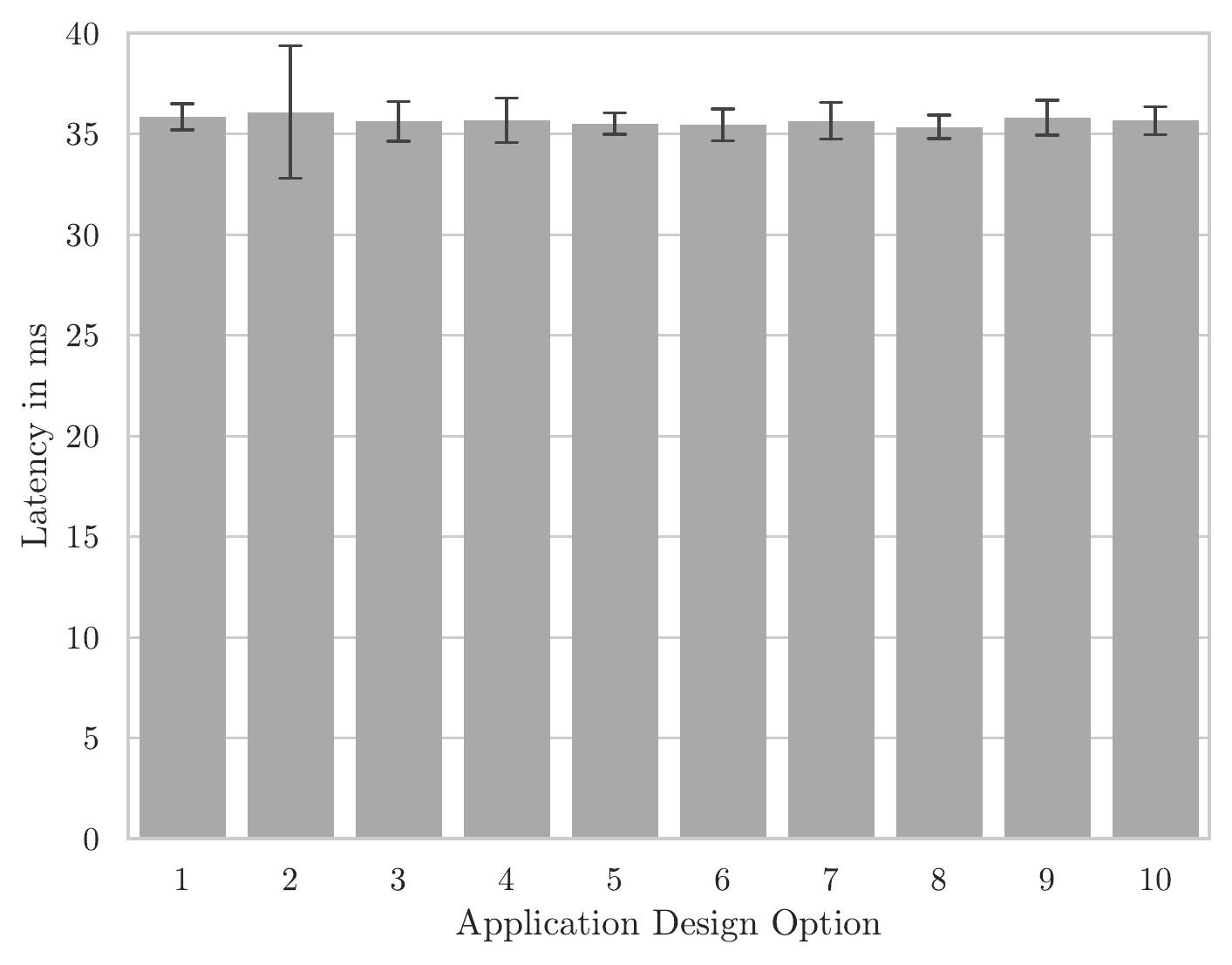}
        \caption{Application Path A1}
        \label{fig:mockfog-a1}
    \end{subfigure}\hfill
    \begin{subfigure}{0.45\textwidth}
        \includegraphics[width=\linewidth]{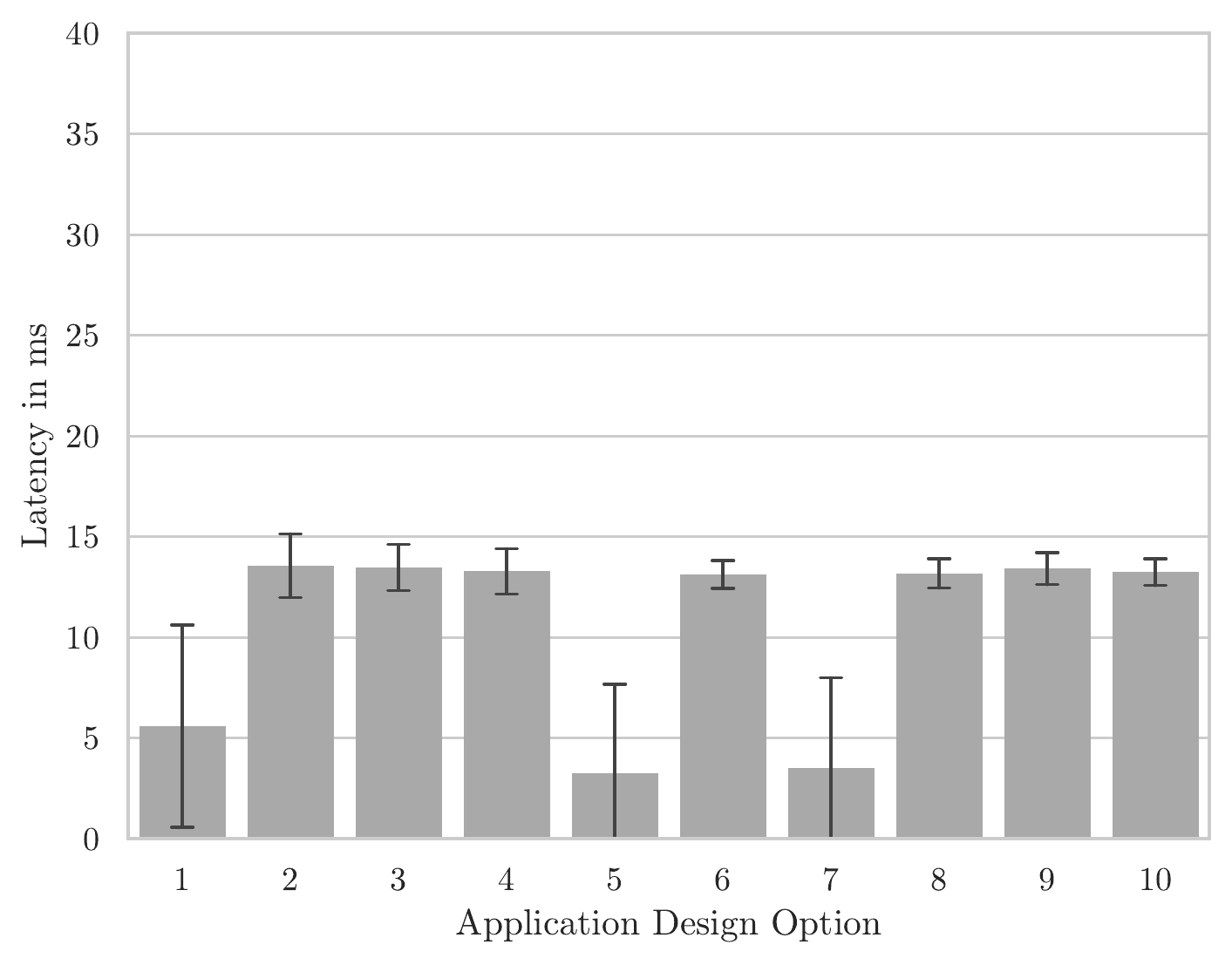}
        \caption{Application Path A2}
        \label{fig:mockfog-a2}
    \end{subfigure}

    \begin{subfigure}{0.45\textwidth}
        \includegraphics[width=\linewidth]{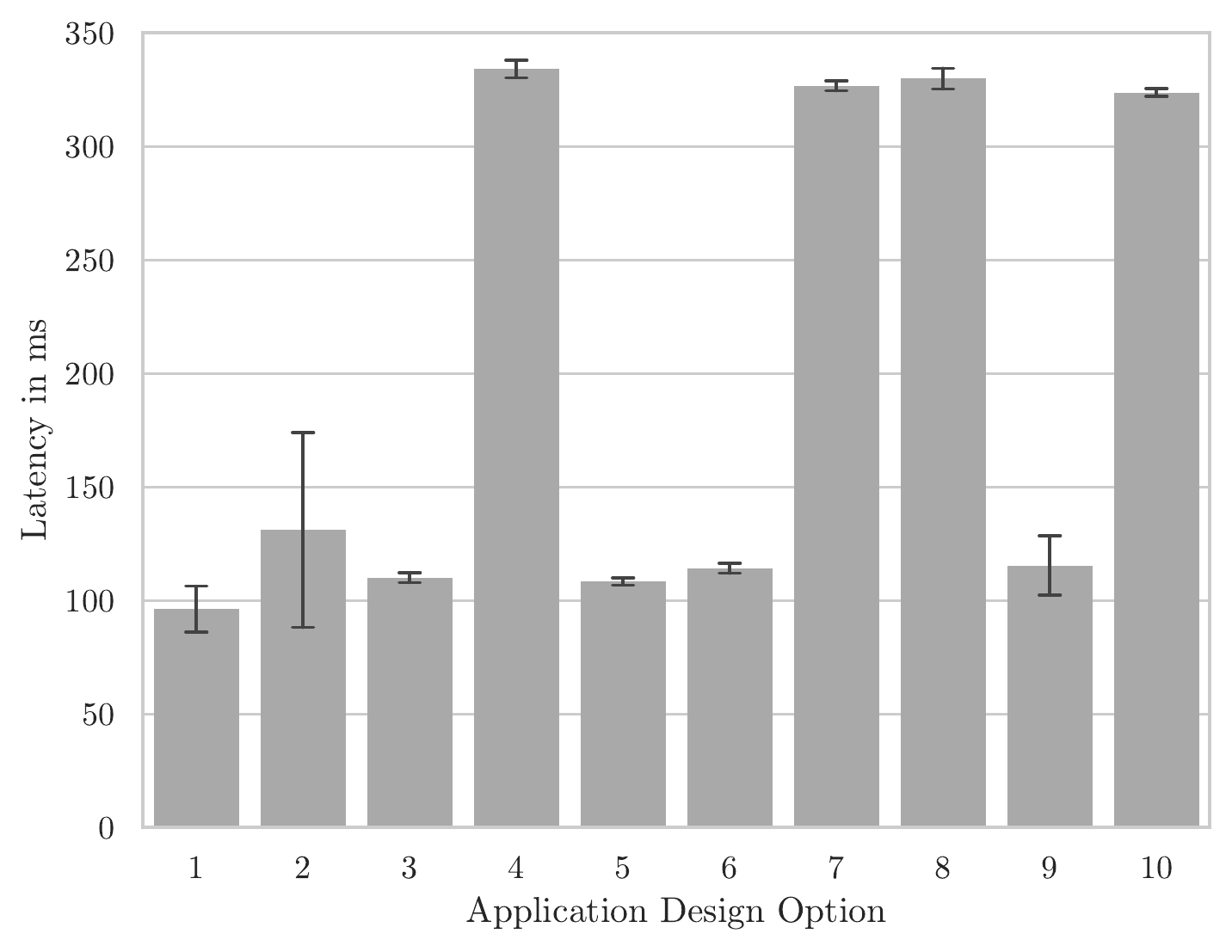}
        \caption{Application Path A3}
        \label{fig:mockfog-a3}
    \end{subfigure}\hfill
    \begin{subfigure}{0.45\textwidth}
        \includegraphics[width=\linewidth]{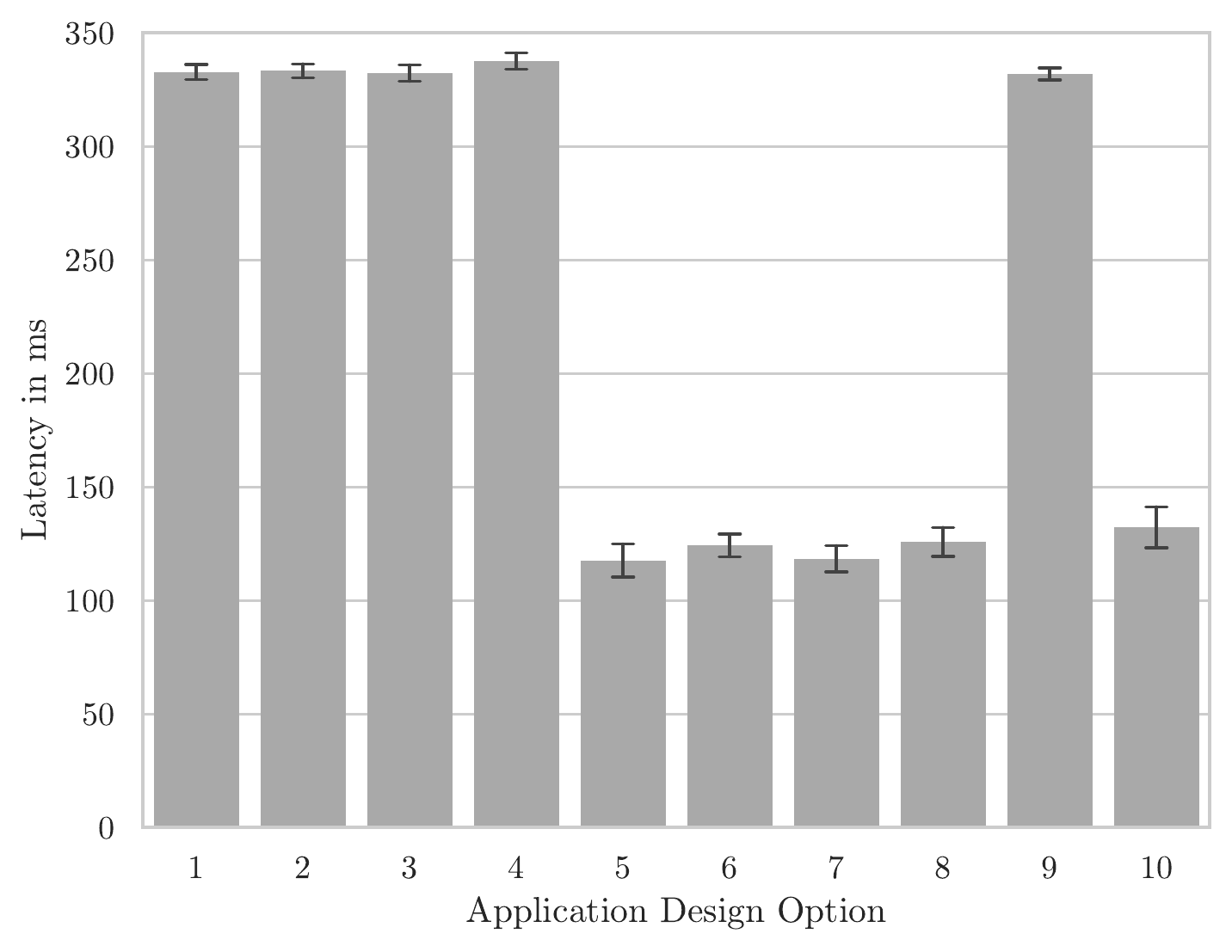}
        \caption{Application Path A4}
        \label{fig:mockfog-a4}
    \end{subfigure}
    \caption{Results for testbed experiments with MockFog. We show average end-to-end latency measured for all application design options for each application path. Error bars show the standard deviation. Application design option \emph{5} consistently is among those with the lowest end-to-end latency for each application path.}
    \label{fig:results}
\end{figure}

MockFog sets artificial network bandwidth and latency limits between machines and deploys our software components to the machines.
The mappings for sinks and sources are identical each time, for instance with the Camera process running on the Camera node.
Service mappings follow the ten most efficient design options identified through simulation.
For each option, MockFog runs the application for 20 minutes and then collects logs to determine end-to-end latency for each application path.
We repeat this process three times to gain accurate results and use median results in further analysis.

We measure end-to-end latency by attaching timestamps and unique identifiers to each request that passes through the system.
Each component logs when it sends or receives a request with a specific identifier.
One problem with measuring end-to-end latency in this manner is clock skew.
When the clocks of two machines are not in sync, the measurement can become inaccurate.
To limit this effect, all machines synchronize their clocks through the AWS Time Sync Service in their region before the experiments run.
This resulted in clock deviations of under 0.3ms during our experiments.

Between re-runs of the same experiment setup, we see a small overall coefficient of variation of between 0\% and 3\%.
Consequently, we can say that our experiment results are robust.
We use the average end-to-end latency unless stated otherwise and show these results in Figure~\ref{fig:results}.
As expected, latency for the A1 application path is similar across all design options, as the Check for Defects service is always deployed to the same kind of Factory Data Center.
On the A2 application path, we observe an end-to-end latency of between 3ms and 4ms when the Adapt service runs on the Packaging Controller and 14ms when placed on the Factory Data Center, due to the increase in network latency caused by additional hops for each request.
This difference is even greater when considering only the Sensor source, where end-to-end latency is under a millisecond when the Adapt service is deployed on the Packaging Controller.
For the A3 application path, processing latency of the Predict service is higher when it runs on the Factory Data Center, with an average latency of 89ms for application design option 1, and even higher for options 2 and 9, where the Check for Defects, Adapt Machine, and Predict service are all deployed on this node, with 123ms and 108ms, respectively.
When the Predict service runs on the Office Data Center or Cloud, this processing latency is lower, between 67ms and 77ms.
For placement on the Cloud node, this reduction of processing latency is offset by a considerable increase in network latency to 257ms.
The Aggregate service of application path A4 has a processing latency of between 0.1ms and 0.15ms, regardless of the machine type of the Wireless Gateway, to which this service is always deployed.
At this scale, this difference could also be attributed to measurement error.
The Generate Dashboard service has a lower processing latency when deployed to the Cloud at 89ms to 90ms than when deployed to the Factory Data Center, where processing latency ranges from 95ms up to 109ms.
Yet again this difference is offset by transmission latency, which, is lower here at 23ms as compared to 243ms.

As already ensured through simulation with FogExplorer, all application design options we benchmarked on the MockFog testbed comply with all SLOs defined for the application paths.

\subsubsection{Determining the Final Application Design}
\label{subsubsec:casestudyresults}

\begin{figure}
    \centering
    \includegraphics[width=\textwidth]{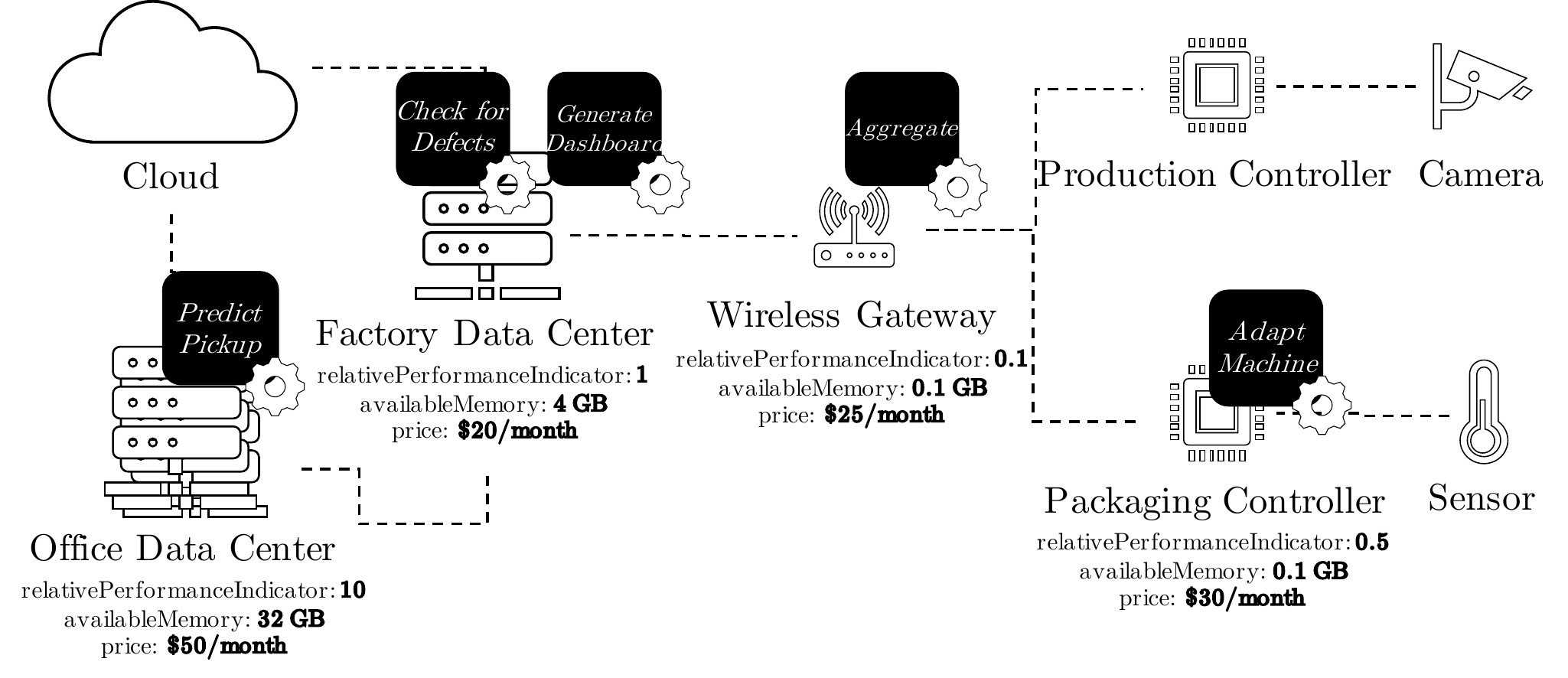}
    \caption{Service mapping and infrastructure option in the best application design option as determined in our case study}
    \label{fig:final}
\end{figure}

Using the results from our MockFog experiments, we can now discard more application design options.
Of the ten application design options we deployed to the emulated testbed, option \emph{5} is the most efficient.
We show service mapping and determined infrastructure options in Figure~\ref{fig:final}.
Here, the Factory Data Center hosts the Check for Defects and Generate Dashboard services, the Adapt Machine service is placed on the Packaging Controller, the Predict Pickup service on the Office Data Center, and the Wireless Gateway is used for the Aggregate service.
As infrastructure options, we use the smallest available machines for the Wireless Gateway and Office Data Center, and the medium option for the Factory Data Center.
In this application design option, the Cloud is not used to host any services, hence we do not require a machine there.
Here, we skip the optional deployment of several options on a physical fog testbed as described in Section~\ref{sec:resolution} since we will do exactly that in our evaluation of result quality in Section~\ref{subsec:results}.

\subsection{Result Evaluation}
\label{subsec:results}

After having shown the applicability of our process through a case study, we now evaluate it by deploying our resulting architecture on a physical testbed.
We benchmark our application with a synthetic workload and determine whether our process has really converged toward the most efficient design by comparing it to application design options that were discarded in earlier steps of the process.

\begin{table}
    \renewcommand{\arraystretch}{1.3}
    \caption{Overview of placement options and the step in which the option was discarded. This shows that early process steps alone cannot provide good enough recommendations.}
    \label{tab:matrix}
    \centering
    \resizebox{\textwidth}{!}{
        \begin{tabular}{c c c c c c c c c c}
                                                                         &                            & Camera                                          & \begin{tabular}[c]{@{}c@{}}Production\\Controller\end{tabular}                       & Sensor                                          & \begin{tabular}[c]{@{}c@{}}Packaging\\Controller\end{tabular}                                      & \begin{tabular}[c]{@{}c@{}}Wireless\\Gateway\end{tabular}                                      & \begin{tabular}[c]{@{}c@{}}Factory\\Data Center\end{tabular}                                      & \begin{tabular}[c]{@{}c@{}}Office\\Data Center\end{tabular}                                         & Cloud                                                             \\
            \rotatebox[origin=c]{90}{\textbf{A1}}                        & \begin{tabular}[c]{@{}c@{}}Check for\\Defects\end{tabular}  & \cellcolor{gray}Simulation                      & \cellcolor{gray}Simulation                      & \cellcolor{lightgray}\begin{tabular}[c]{@{}c@{}}Best\\Practices\end{tabular}  & \cellcolor{lightgray}\begin{tabular}[c]{@{}c@{}}Best\\Practices\end{tabular}                 & \cellcolor{gray}Simulation                                     & \textcolor{white}{\cellcolor{black}\begin{tabular}[c]{@{}c@{}}Final\\Design\end{tabular}}  & \cellcolor{lightgray}\begin{tabular}[c]{@{}c@{}}Best\\Practices\end{tabular}                   & \cellcolor{lightgray}\begin{tabular}[c]{@{}c@{}}Best\\Practices\end{tabular}                   \\ \cline{1-2}
            \rotatebox[origin=c]{90}{\textbf{A2}}                        & \begin{tabular}[c]{@{}c@{}}Adapt\\Machine\end{tabular} & \cellcolor{lightgray}\begin{tabular}[c]{@{}c@{}}Best\\Practices\end{tabular} & \cellcolor{lightgray}\begin{tabular}[c]{@{}c@{}}Best\\Practices\end{tabular} & \cellcolor{gray}Simulation                      & \textcolor{white}{\cellcolor{black}\begin{tabular}[c]{@{}c@{}}Final\\Design\end{tabular}} & \cellcolor{gray}Simulation                                     & \cellcolor{gray}Simulation                                     & \cellcolor{lightgray}\begin{tabular}[c]{@{}c@{}}Best\\Practices\end{tabular}                   & \cellcolor{lightgray}\begin{tabular}[c]{@{}c@{}}Best\\Practices\end{tabular}                   \\ \cline{1-2}
            \rotatebox[origin=c]{90}{\textbf{A3}}                        & \begin{tabular}[c]{@{}c@{}}Predict\\Pickup\end{tabular} & \cellcolor{lightgray}\begin{tabular}[c]{@{}c@{}}Best\\Practices\end{tabular} & \cellcolor{lightgray}\begin{tabular}[c]{@{}c@{}}Best\\Practices\end{tabular} & \cellcolor{lightgray}\begin{tabular}[c]{@{}c@{}}Best\\Practices\end{tabular} & \cellcolor{lightgray}\begin{tabular}[c]{@{}c@{}}Best\\Practices\end{tabular}                & \cellcolor{lightgray}\begin{tabular}[c]{@{}c@{}}Best\\Practices\end{tabular}                & \textcolor{white}{\cellcolor{black}\begin{tabular}[c]{@{}c@{}}Final\\Design\end{tabular}} & \textcolor{white}{\cellcolor{darkgray}\begin{tabular}[c]{@{}c@{}}Emulated\\Testbed\end{tabular}} & \textcolor{white}{\cellcolor{darkgray}\begin{tabular}[c]{@{}c@{}}Emulated\\Testbed\end{tabular}} \\ \cline{1-2}
            \multirow{2}{*}[-2ex]{\rotatebox[origin=c]{90}{\textbf{A4}}} & Aggregate                  & \cellcolor{gray}Simulation                      & \cellcolor{gray}Simulation                      & \cellcolor{gray}Simulation                      & \cellcolor{gray}Simulation                                     & \textcolor{white}{\cellcolor{black}\begin{tabular}[c]{@{}c@{}}Final\\Design\end{tabular}} & \cellcolor{gray}Simulation                                     & \cellcolor{lightgray}\begin{tabular}[c]{@{}c@{}}Best\\Practices\end{tabular}                   & \cellcolor{lightgray}\begin{tabular}[c]{@{}c@{}}Best\\Practices\end{tabular}                   \\ \cline{2-2}
                                                                         & \begin{tabular}[c]{@{}c@{}}Generate\\Dashboard\end{tabular} & \cellcolor{lightgray}\begin{tabular}[c]{@{}c@{}}Best\\Practices\end{tabular} & \cellcolor{lightgray}\begin{tabular}[c]{@{}c@{}}Best\\Practices\end{tabular} & \cellcolor{lightgray}\begin{tabular}[c]{@{}c@{}}Best\\Practices\end{tabular} & \cellcolor{lightgray}\begin{tabular}[c]{@{}c@{}}Best\\Practices\end{tabular}                & \cellcolor{lightgray}\begin{tabular}[c]{@{}c@{}}Best\\Practices\end{tabular}                & \textcolor{white}{\cellcolor{black}\begin{tabular}[c]{@{}c@{}}Final\\Design\end{tabular}} & \cellcolor{gray}Simulation                                        & \textcolor{white}{\cellcolor{darkgray}\begin{tabular}[c]{@{}c@{}}Emulated\\Testbed\end{tabular}} \\
        \end{tabular}
    }
\end{table}
\begin{figure}
    \centering
    \begin{subfigure}{0.45\textwidth}
        \includegraphics[width=\linewidth]{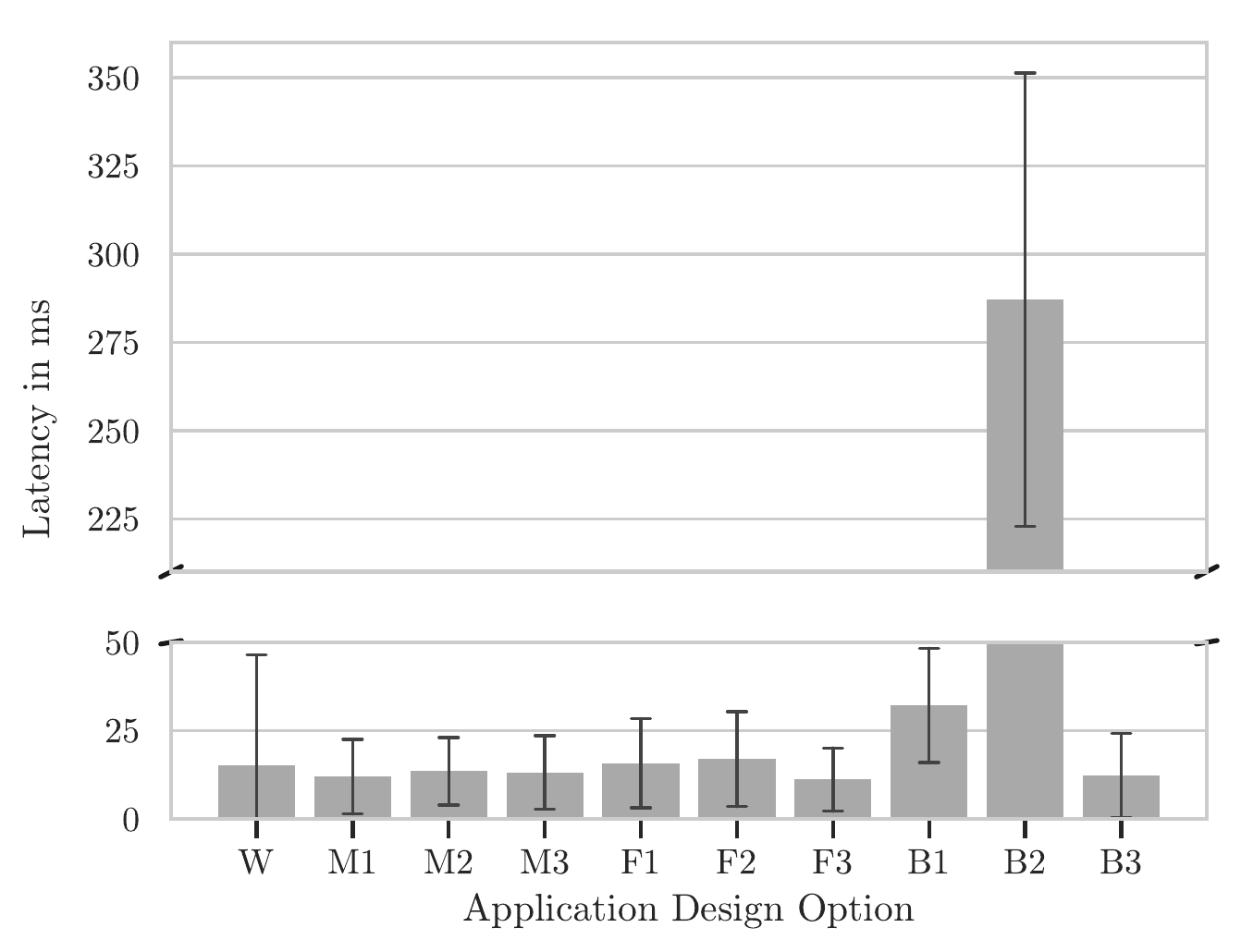}
        \caption{Application Path A1}
        \label{fig:real-a1}
    \end{subfigure}\hfill
    \begin{subfigure}{0.45\textwidth}
        \includegraphics[width=\linewidth]{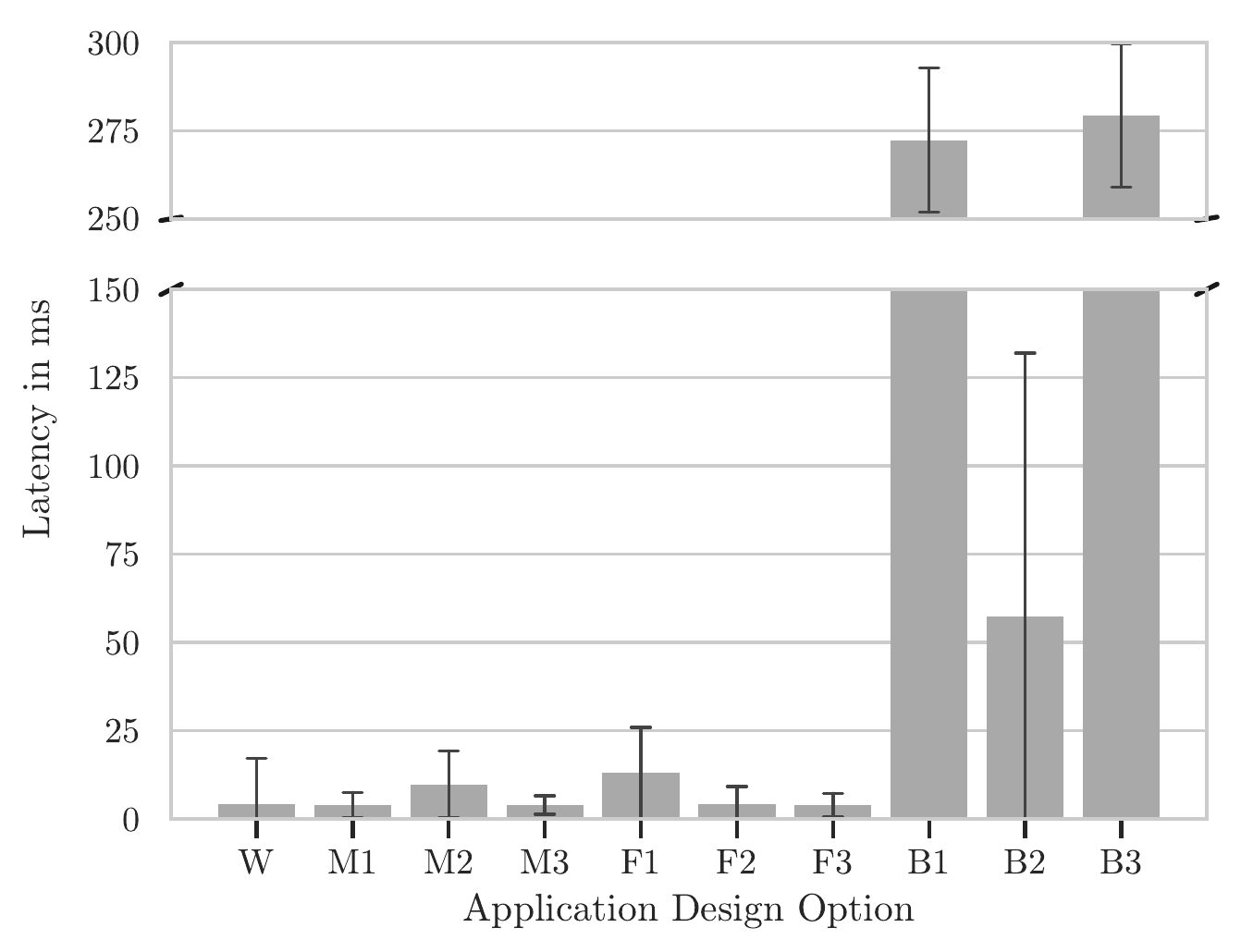}
        \caption{Application Path A2}
        \label{fig:real-a2}
    \end{subfigure}

    \begin{subfigure}{0.45\textwidth}
        \includegraphics[width=\linewidth]{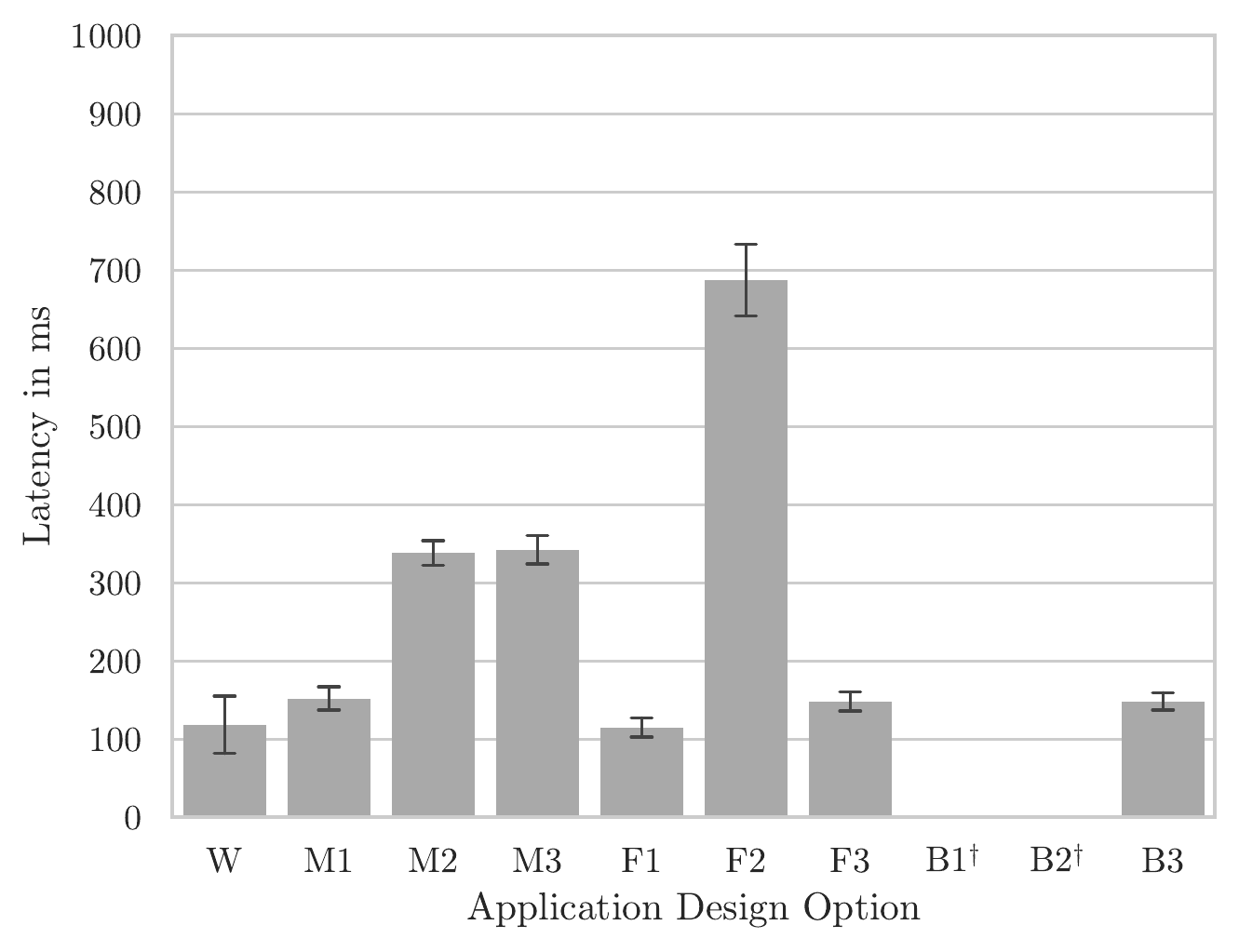}
        \caption{Application Path A3}
        \label{fig:real-a3}
    \end{subfigure}\hfill
    \begin{subfigure}{0.45\textwidth}
        \includegraphics[width=\linewidth]{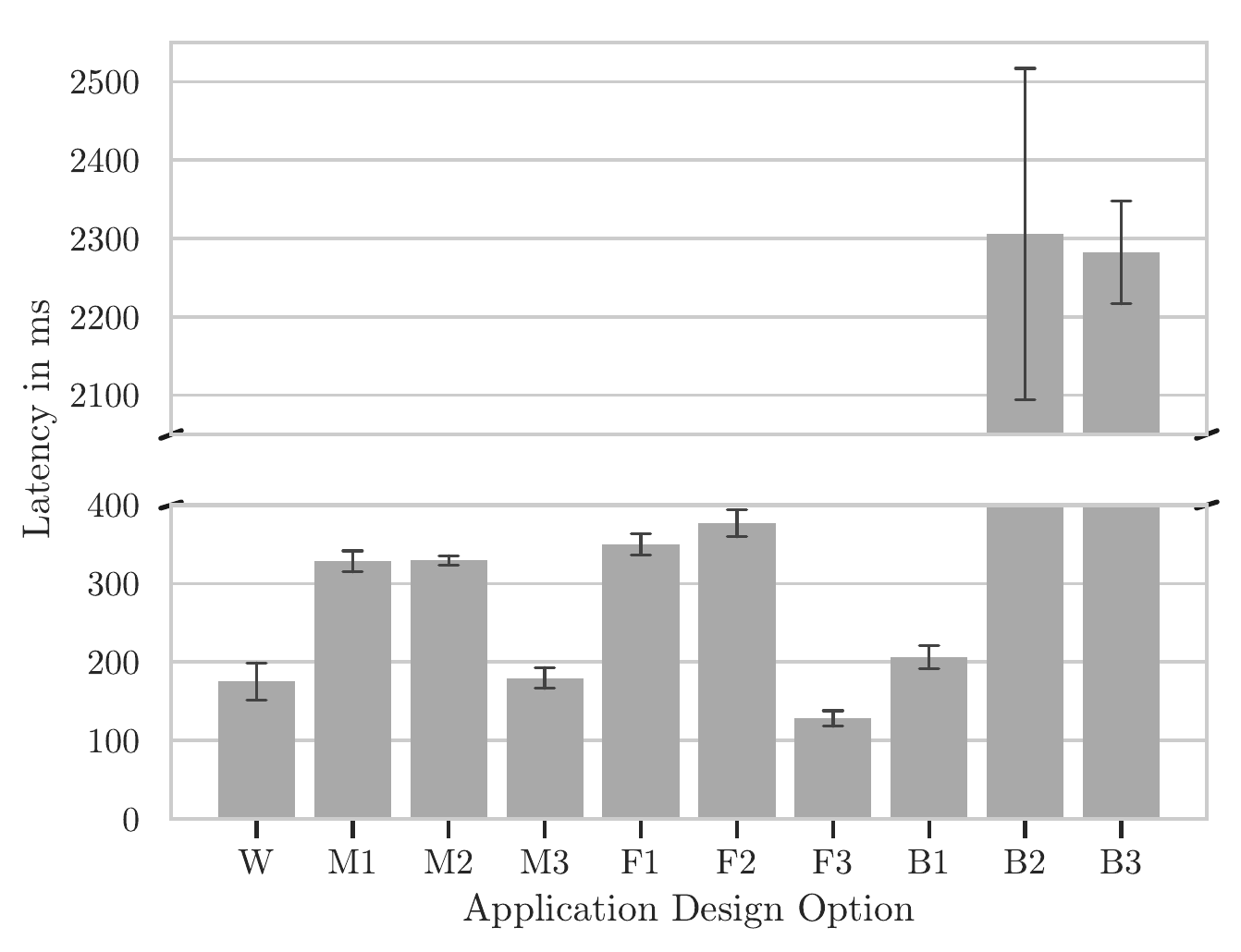}
        \caption{Application Path A4}
        \label{fig:real-a4}
    \end{subfigure}
    \caption{Latency results for experiments on the physical testbed. We show average end-to-end latency measured for all application design options for each application path. Error bars show the standard deviation.\\
        \textsuperscript{†} Application design options B1 and B2 were unable to run the Predict Pickup service as the infrastructure component would run out of memory, hence no results for the A3 application path can be shown here.}
    \label{fig:realresults}
\end{figure}

In Table~\ref{tab:matrix} we show application design options and the step in which we filtered them out.
This figure also shows the final application design that our process determined to be the most efficient.
The final design has passed the check for best practices, simulation with FogExplorer, and benchmarking on the emulated MockFog testbed.
We now further evaluate this design by comparing it to other design options that we filtered out during the process.
Obviously, we cannot compare all possible design options.
For each filter we applied, we randomly chose three of the discarded design options, deployed them on a physical testbed and benchmarked them.
\emph{M1-3}, \emph{F1-3}, and \emph{B1-3} denote the three designs that were filtered out by \textbf{M}ockFog, \textbf{F}ogExplorer, and the application of \textbf{b}est practices, respectively.
For sake of comparison, we also deploy and benchmark our final, \textbf{w}inning design as presented in Section~\ref{subsubsec:casestudyresults}, which we denote as \emph{W}.
Software components use the same implementation and deployment method (i.e., Docker containers) as in our emulated MockFog testbed.
Our testbed comprises two \emph{Raspberry Pi 3B+} single-board computers, one acting as Camera and Production Controller, and the other as Sensor and Packaging Controller.
These boards connect over 2.4GHz WiFi to a \emph{MacBook Pro} with an \emph{Intel Core 2 Duo} processor that we use as our Wireless Gateway.
This computer, in turn, connects to a LAN over Gigabit Ethernet.
This network has a 50Mbit/s Internet uplink and a \emph{ThinkPad x220} laptop with an \emph{Intel Core i5} processor that acts as the Factory Data Center connected to it.
Finally, as our Office Data Center, we use a virtual machine instance on AWS EC2 in the \emph{eu-west-1} Ireland region.
As the Cloud instance, we use an AWS EC2 virtual machine instance in the \emph{ap-northeast-2} Tokyo region.
The respective instance types depend on the machine type used in the selected application design, see Table~\ref{tab:instance_types}.
Experiments run for 20 minutes after an initial startup time of 5 minutes and are repeated three times.
We report the results of the median run.
Variance across runs with the same experiment setup was between 1\% and 4\% for all experiments except for setup M3 (9\%) where one outlier had a higher end-to-end latency for the A3 application path, and experiments B1 (15\%) and B3 (6\%) that were unable to complete correctly.

Figure~\ref{fig:realresults} shows the average transmission and processing times measured in our experiments.
Experiments for application design options B1 and B2 were unable to complete as the Predict Pickup service ran out of memory on the Packaging Controller and Wireless Gateway, respectively, where it was deployed with these design options.
The B3 option, while able to run all services, leads to a higher latency than others that were selected with the first step of our process.
Design option F1 was determined by FogExplorer to comply with all SLOs, yet was not in the 95\textsuperscript{th} percentile cost-wise and was hence discarded.
Nevertheless, latency measurements appear to be on par with designs W and M1 through M3.
Option F2 violates SLO requirements in the simulation and we observe that it is also less efficient than others we tested, so this elimination was correct.
Finally, while FogExplorer discards F3 for insufficient resources, as the Wireless Gateway component here has too little available memory for the Check for Defects service, we were able to deploy it correctly on our physical testbed and latency is similar to our winning design option W.
Yet this deployment is more costly than W as it uses more expensive infrastructure components.
For options W and M1 through M3, we see results as in MockFog where we tested these design options already.
Consequently, design option W again is the most efficient option among those.

\section{Discussion and Limitations}
\label{sec:discussion}

The five-step design process we propose can help to address the challenge of designing efficient fog-based IoT applications.
Yet as with all tools, it is important to know its limits to employ it correctly.
First and foremost, our proposed process targets static applications.
Although not all information about the system is necessarily required upfront and infrastructure and software models are extended and modified along the way, as we have described, our design process is not equipped to deal with dynamic deployment changes such as would be necessary for physically moving sources, sinks, or compute nodes.
For example, in order to augment the application with a new service, parts of the process would need to be re-run from the start.
While simulation and testbed emulation can be automated, best practices would need to be applied by an actual application design engineer.

While networks with mobile nodes, frequent outages, or regular changes in topology may exist, we envision that static applications such as the smart factory in our case study are common.
Furthermore, our process can be used for the static components of a more dynamic application while the dynamic components are deployed using other approaches such as~\cite{paper_bermbach_auctions4function_placement}.

Additionally, we want to emphasize again that our process is an \emph{offline} approach, i.e., it is detached from the actual deployed application.
Conversely, an \emph{online} approach to designing fog-based IoT applications would interact with the deployed application and infrastructure to collect metrics or logs and could then move application components around, possibly even making modifications to the infrastructure.
An online approach has the benefit that it requires less upfront research and investment and that it can also support dynamically changing applications to some extent.
Nevertheless, we find that an offline approach has some key advantages:
First, it does not interfere with the production application, as no additional monitoring or orchestration components degrade application QoS.
Second, it facilitates infrastructure planning alongside application development.
Rather than relying on on-demand fog infrastructure, which might not be easily available, our offline process aids in determining the optimal infrastructure components and their sizing.
And third, only a process with a human in the loop can benefit from domain knowledge that is not easily quantifiable.
Application developers have the option to step in and adjust the outputs of each step of the process as they see fit, which also helps provide more understandable results that can be considered.
The human involvement during the individual steps depend largely on the selected tooling.
For instance, MockFog experiments can be conducted in a completely automated fashion.

Another challenge is the number of factors at play in fog application design.
We quantify the features of application and infrastructure components.
Availability, performance, network latency, or available network bandwidth may be subject to external influence factors. however.
For example, sharing a network connection with a different tenant in a cloud data center, an application service executing slower for certain inputs, energy consumption, or job backlogs through failing components can all influence availability, latency, or cost as well.
Abstracting from such factors in our models means that our simulation and testbed experiments cannot accurately reflect results that we would observe in the real world.
We argue, however, that we need this abstraction to keep models and simulation simple, which is in turn necessary to facilitate their use in such a design process.
These factors can then be tested later in the process using physical testbeds.

In Section~\ref{sec:fogexplorer}, we introduced SLOs for application paths as a way to convert the multi-objective optimization of cost and service latency for each path into a single-objective optimization of cost within the specified latency constraints.
While reducing end-to-end latency is always better, we argue that additional investment can lead to diminishing returns after a certain point.
Finding these fixed constraints, however, can be difficult for system designers and setting SLOs too low or too high can have negative impacts on the overall satisfaction with the final application design by unnecessarily increasing cost or latency, respectively.
In future work, we want to further explore this relationship between cost and utility of reduced latency so that this decision can be made on a more informed basis.

\section{Related Work}
\label{sec:relatedwork}

We have described how the correct placement of IoT application components in the fog is difficult yet crucial for an efficient use of resources.
This is a known research problem and has been discussed in existing publications.
Below we provide an overview of existing approaches for fog application design and indicate how they compare to our approach in Table~\ref{tab:fogapplication_relwork}.

\begin{table}[!t]
    \renewcommand{\arraystretch}{1}

    \caption{Ease and accuracy trade-off in state-of-the-art approaches to fog application design. To quantify ease of use, we extrapolate the time for a complete investigation of all application design options from our experience with our case study. We also leverage that experience to quantify an estimated reduction of the total solution space as a metric for the accuracy of different approaches. By combining different approaches into one process, we can increase result accuracy without sacrificing ease of use.}
    \label{tab:fogapplication_relwork}
    \centering
    \resizebox{0.9\textwidth}{!}{

        \begin{tabular}{c|c|c|c}
            \diagbox{\begin{tabular}[c]{@{}c@{}}\textbf{Ease}\\\textbf{of Use}\end{tabular}}{\textbf{Accuracy}} & \begin{tabular}[c]{@{}c@{}}\textbf{Low}\\{\footnotesize Reduction of solution space: 30-50\%}\end{tabular}  & \begin{tabular}[c]{@{}c@{}}\textbf{Medium}\\{\footnotesize Reduction of solution space: 50-80\%}\end{tabular}  & \begin{tabular}[c]{@{}c@{}}\textbf{High}\\{\footnotesize Reduction of solution space: 80-99\%}\end{tabular}  \\ \hline
            \begin{tabular}[c]{@{}c@{}}\textbf{Low}\\{\footnotesize Est. time: weeks}\end{tabular}                              &                            &                            & \begin{tabular}[c]{@{}c@{}}Emulation with Benchmarks~\cite{Eisele2017-yx,Mayer2017-dt,Hasenburg2019-er,Coutinho2018-wa}\\{\footnotesize Time: 4 weeks}\\{\footnotesize Reduction of solution space: 99\%}\end{tabular}  \\ \hline
            \begin{tabular}[c]{@{}c@{}}\textbf{Medium}\\{\footnotesize Time: Days}\end{tabular}                              &                            & \begin{tabular}[c]{@{}c@{}}Simulation~\cite{Hasenburg2018-fn,Hasenburg2018-nd,Gupta2017-jx,Qayyum2018-ml}\\{\footnotesize Time: 1 week}\\{\footnotesize Reduction of solution space: 80\%}\end{tabular}  &                            \\ \hline
            \begin{tabular}[c]{@{}c@{}}\textbf{High}\\{\footnotesize Time: Hours}\end{tabular}                              & \begin{tabular}[c]{@{}c@{}}Best Practices and Reference Architectures~\cite{paper_pfandzelter_streams_functions,Santos2020-qx}\\{\footnotesize Time: 1 hour}\\{\footnotesize Reduction of solution space: 50\%}\end{tabular} & \begin{tabular}[c]{@{}c@{}}Formal Optimization Models~\cite{Shekhar2020-yg,Heintz2020-lo,Brogi2017-fk,Brogi2017-nl,Skarlat2017-ye}\\{\footnotesize Est. time: 1 day}\\{\footnotesize Est. reduction of solution space: 60\%}\end{tabular} & \begin{tabular}[c]{@{}c@{}}\emph{Zero to Fog}\\{\footnotesize Time: 2 days}\\{\footnotesize Reduction of solution space: 99\%}\end{tabular} \\ \hline
        \end{tabular}
    }
\end{table}

Brogi et al.~\cite{Brogi2017-fk} present \emph{FogTorch}, which models fog infrastructure by parameterizing available fog nodes, communication links, end devices, application components, and QoS constraints, and then finds eligible deployments of application components.
While this approach leads to a set of valid application deployment options, solving fog application deployment in this manner is NP-hard, as the authors show.
Consequently, finding valid deployment options becomes exponentially harder with each added component and is infeasible for larger deployments.
Tong et al.~\cite{Tong2016-ke} and, to some extent Heintz et al.~\cite{Heintz2020-lo} take a similar approach to FogTorch, while~\cite{Skarlat2017-ye,Xu2019-rq,Mahmud2018-ns,Hong2016-hv,Brogi2017-nl,Cardellini2016-gw,Shekhar2020-yg,Oh2020-xj,8975987,9132684,9305277,9339982,9358007,8955944,9373980,9361310,8869806,8960404,8897679,spe.2896,spe.2951,spe.2942} employ a more efficient heuristics approach to solve the formalized optimization problem.
Naas et al.~\cite{Naas2017-ln} offer a heuristics-based solution as well, but place data replicas rather than services, a challenge that is also addressed in~\cite{Hasenburg2020-yo,Hasenburg2019-oe}.
Formulating such assignment problems requires complete information about the system upfront, including infrastructure and software implementation details.
This may be available for existing applications that are moved to a fog infrastructure, yet allows little room for flexibility.
For agile development of new applications, however, these details are only slowly emerging.
Inarguably, these approaches find optimal solutions in static analysis, yet we propose that benchmarks on emulated or physical testbeds are necessary to verify that calculated results hold up in a real deployment.

Khare et al.~\cite{Khare2019-ef} also employ heuristics to create an efficient application design for distributed, edge-based stream processing.
Additionally, they also employ them in a multi-step process, where a DAG of the entire application is first split into a set of linear chains for which latency is estimated individually, similar to the application paths we introduced in Section \ref{sec:models}.
The authors here, however, approximate these processing chains algorithmically, which is an interesting alternative approach as it leads to less overhead for application designers, albeit by sacrificing accuracy.

\emph{Fogernetes} as proposed in~\cite{Wobker2018-ud} automates the deployment of software services across a number of fog nodes by leveraging the Kubernetes orchestration, as Santos et al.~\cite{Santos2019-en} have also proposed.
Similarly,~\cite{Saurez2016-oo,Santoro2017-vl,Skarlat2018-pl} have also presented such dynamic middleware.
While these systems are flexible, they can only optimize latency and do not take system costs into account.
Rather, they assume that a specific set of infrastructure already exists along with a mapping that does not lead to under-provisioning.
In our proposed process, we provision only infrastructure that is really needed, keeping overall cost to a minimum.
We argue that a more efficient fog application can be designed by building the underlying infrastructure in parallel.
Furthermore, the infrastructure is often not yet fixed at the start of the development process.

To this end, Roy et al.~\cite{Roy2011-uy} present MAQ-PRO, a process for infrastructure capacity planning for component-based applications that is similar to our proposed process.
MAQ-PRO begins with a profile of components, analysis of the application scenario (cf. Section~\ref{sec:models}), and a base performance model (cf. Section~\ref{sec:fogexplorer}), and it also considers SLA bounds and workloads.
Their approach, however, is unsuitable for the novel paradigm of fog computing as it does not consider network distance between infrastructure components, which is crucial in the fog.

In Section~\ref{sec:fogexplorer} we propose using FogExplorer to simulate fog placement.
Alternatively, Gupta et al.~\cite{Gupta2017-jx} have proposed the \emph{iFogSim} tool to model and simulate the use of fog application resources.
Their tool, however, has constraints in that it only allows tree-shaped infrastructure models, which is not representative of most fog infrastructure that can contain cycles, such as in our case study.
Furthermore, their tool requires highly detailed application traces, which are not feasible this early in the design phase.
A further alternative is \emph{IoTSim-Edge}~\cite{spe.2787}, a simulation framework for IoT application in fog environments based on \emph{CloudSim}.
As our process only defines abstract steps, both simulation tools could be used in our process instead of FogExplorer if the user so wishes.

In~\cite{Brambilla2014-kl}, Brambilla et al.~present an approach for simulating large scale sensor networks for the IoT.
While useful in its own right, it lacks an estimation of system cost.
We target more heterogeneous fog networks, albeit at a lower scale.
Additionally,~\cite{Sotiriadis2014-uz,Zeng2017-lq,Qayyum2018-ml,Fernandez-Cerero2020-oa,Sonmez2018-sq,Giang2015-ws} also present simulation tools that could be applied to fog computing.

We also propose using MockFog as an emulated testbed for different application designs in Section~\ref{sec:mockfog}.
Besides MockFog, other application testbeds exist as well.
Eisele et al.~\cite{Eisele2017-yx} propose a hardware-in-the-loop simulation that uses a simulation tool in conjunction with a physical testbed.
This allows them to leverage flexibility in workload generation from the simulation tool but a realistic environment from the physical testbed.
However, it also leads to increased cost without being entirely accurate.
The \emph{D-Cloud}~\cite{Banzai2010-lr} software testing framework allows individual software components to be placed on different virtual machines to emulate a cloud environment.
This tool, however, cannot be applied to a fog infrastructure.
Furthermore, Coutinho et al.~\cite{Coutinho2018-wa}, and Mayer et al.~\cite{Mayer2017-dt} propose \emph{Fogbed} and \emph{EmuFog}, which use the network simulators Mininet and Maxinet~\cite{De_Oliveira2014-bq}, to test distributed fog applications.
Yet unlike MockFog, these testbeds can only simulate realistic network conditions, not the constrained compute capabilities of fog nodes, especially at the edge.
Balasubramanian et al.~\cite{Balasubramanian2014-tj} present a testbed for fog applications that facilitates emulating these constraints but requires physical hardware for each node rather than cheaper virtual machines.

Luthra et al.~\cite{Luthra2019-uh} present \emph{ProgCEP}, an operator-based programming model for complex event processing in the fog.
ProgCEP allows placing application operators on fog nodes through an API and facilitates QoS monitoring of that application.
While the system does not offer its own operator placement algorithm, it aims to aid the development of algorithms.
It could therefore be used to test different application design options on physical infrastructure as is part of our design process.
A similar dynamic migration is also implemented in \emph{FogBus}~\cite{Tuli2019-xs}.

To the best of our knowledge, our work is the first that combines best practices, simulation, and emulation into a complete design process for fog-based IoT applications.

\section{Conclusion}
\label{sec:conclusion}
Engineering IoT applications in an efficient way is challenging as the process needs to consider both software architecture and its deployment to a physical infrastructure.
Existing approaches can only provide limited guidance since they are either based on theoretical models and simulation (i.e., inherently limited in their accuracy) or based on experiment testbeds (i.e., the evaluation effort is too high to explore more than a few design options).

In this paper, we have proposed a five-step process for designing efficient fog-based IoT applications that integrates and extends previous work of ours.
Rather than relying solely on global optimization, simulation, or testbed benchmarking, we combine best practices, simulation, and testbed evaluation to choose the most efficient infrastructure options and software service placements from an exponentially growing pool of deployment options.
Furthermore, we have shown the effectiveness of this approach through a smart factory case study.
By deploying different options on a physical testbed, we also showed that our process identified an efficient application design in our case study and, by extension, that our process achieves the desired results.

\bibliographystyle{IEEEtran}
\bibliography{bibliography}

\begin{thebibliography}{10}
\providecommand{\url}[1]{#1}
\csname url@samestyle\endcsname
\providecommand{\newblock}{\relax}
\providecommand{\bibinfo}[2]{#2}
\providecommand{\BIBentrySTDinterwordspacing}{\spaceskip=0pt\relax}
\providecommand{\BIBentryALTinterwordstretchfactor}{4}
\providecommand{\BIBentryALTinterwordspacing}{\spaceskip=\fontdimen2\font plus
\BIBentryALTinterwordstretchfactor\fontdimen3\font minus
  \fontdimen4\font\relax}
\providecommand{\BIBforeignlanguage}[2]{{%
\expandafter\ifx\csname l@#1\endcsname\relax
\typeout{** WARNING: IEEEtran.bst: No hyphenation pattern has been}%
\typeout{** loaded for the language `#1'. Using the pattern for}%
\typeout{** the default language instead.}%
\else
\language=\csname l@#1\endcsname
\fi
#2}}
\providecommand{\BIBdecl}{\relax}
\BIBdecl

\bibitem{Zhang2015-cb}
B.~Zhang, N.~Mor, J.~Kolb, D.~S. Chan, K.~Lutz, E.~Allman, J.~Wawrzynek,
  E.~Lee, and J.~Kubiatowicz, ``The cloud is not enough: Saving iot from the
  cloud.''\hskip 1em plus 0.5em minus 0.4em\relax 7th {USENIX} Workshop Hot
  Topics in Cloud Comput. ({HotCloud} 15), Jul. 2015, pp. 21--27.

\bibitem{Bermbach2018-bb}
D.~Bermbach, F.~Pallas, D.~G. P{\'e}rez, P.~Plebani, M.~Anderson, R.~Kat, and
  S.~Tai, ``A res. perspective on fog computing.''\hskip 1em plus 0.5em minus
  0.4em\relax {Service-Oriented} Comput. -- {ICSOC} 2017 Workshops, Jun. 2018,
  pp. 198--210.

\bibitem{Bonomi2012-if}
F.~Bonomi, R.~Milito, J.~Zhu, and S.~Addepalli, ``Fog comput. and its role in
  the internet of things.''\hskip 1em plus 0.5em minus 0.4em\relax Proc. 1st
  Ed. MCC Workshop Mobile Cloud Computing, Aug. 2012, pp. 13--16.

\bibitem{Brogi2017-nl}
A.~Brogi, S.~Forti, and A.~Ibrahim, ``How to best deploy your fog applications,
  probably.''\hskip 1em plus 0.5em minus 0.4em\relax 2017 {IEEE} 1st Internat.
  Conf. Fog and Edge Comput. ({ICFEC}), May 2017, pp. 105--114.

\bibitem{Skarlat2017-ye}
O.~Skarlat, M.~Nardelli, S.~Schulte, M.~Borkowski, and P.~Leitner, ``Optimized
  {IoT} service placement in the fog,'' \emph{Service Oriented Comput. and
  Applications}, vol.~11, no.~4, pp. 427--443, 2017.

\bibitem{Mahmud2018-ns}
R.~Mahmud, K.~Ramamohanarao, and R.~Buyya, ``Latency-aware application module
  manage. for fog comput. environments,'' \emph{ACM Trans. Internet Technol.},
  vol.~19, no.~1, pp. 1--21, 2018.

\bibitem{Hong2016-hv}
H.~Hong, P.~Tsai, and C.~Hsu, ``Dynamic module deployment in a fog comput.
  platform.''\hskip 1em plus 0.5em minus 0.4em\relax 2016 18th {Asia-Pacific}
  Netw. Operations and Manage. Symp. ({APNOMS}), Oct. 2016, pp. 1--6.

\bibitem{Skarlat2017-zx}
O.~Skarlat, M.~Nardelli, S.~Schulte, and S.~Dustdar, ``Towards {QoS-Aware} fog
  service placement.''\hskip 1em plus 0.5em minus 0.4em\relax 2017 {IEEE} 1st
  Internat. Conf. Fog and Edge Comput. ({ICFEC}), May 2017, pp. 89--96.

\bibitem{Hasenburg2018-fn}
J.~Hasenburg, S.~Werner, and D.~Bermbach, ``Supporting the eval. of fog-based
  {IoT} appl. during the des. phase.''\hskip 1em plus 0.5em minus 0.4em\relax
  Proc. 5th Workshop Middleware and Appl. for Internet Things, Dec. 2018, pp.
  1--6.

\bibitem{Hasenburg2018-nd}
------, ``Fogexplorer.''\hskip 1em plus 0.5em minus 0.4em\relax Proc. 19th
  Internat. Middleware Conf. (Demos and Posters), Dec. 2018, pp. 1--2.

\bibitem{Gupta2017-jx}
H.~Gupta, A.~Vahid~Dastjerdi, S.~K. Ghosh, and R.~Buyya, ``{iFogSim}: A toolkit
  for model. and simul. of resource manage. techn. in the internet of things,
  edge and fog comput. environments,'' \emph{Softw. Pract. Exp.}, vol.~47,
  no.~9, pp. 1275--1296, 2017.

\bibitem{paper_pfandzelter_streams_functions}
T.~Pfandzelter and D.~Bermbach, ``{IoT} data process. in the fog: Functions,
  streams, or batch processing?''\hskip 1em plus 0.5em minus 0.4em\relax Proc.
  DaMove, Jun. 2019, pp. 201--206.

\bibitem{Gusev2019-ch}
M.~Gusev, B.~Koteska, M.~Kostoska, B.~Jakimovski, S.~Dustdar, O.~Scekic,
  T.~Rausch, S.~Nastic, S.~Ristov, and T.~Fahringer, ``A deviceless edge
  comput. approach for streaming {IoT} applications,'' \emph{IEEE Internet
  Computing}, vol.~23, no.~1, pp. 37--45, 2019.

\bibitem{Karagiannis2020-kx}
V.~Karagiannis and S.~Schulte, ``Comparison of alternative architectures in fog
  computing.''\hskip 1em plus 0.5em minus 0.4em\relax 2020 {IEEE} 4th Int.
  Conf. Fog and Edge Comput. ({ICFEC}), May 2020, pp. 19--28.

\bibitem{Santos2020-qx}
L.~Santos, E.~Silva, T.~Batista, E.~Cavalcante, J.~Leite, and F.~Oquendo, ``An
  architectural style for internet of things systems.''\hskip 1em plus 0.5em
  minus 0.4em\relax Proc. 35th Annu. {ACM} Symp. Appl. Computing, Mar. 2020,
  pp. 1488--1497.

\bibitem{Hasenburg2019-er}
J.~Hasenburg, M.~Grambow, E.~Gr{\"u}newald, S.~Huk, and D.~Bermbach,
  ``{MockFog}: Emulating fog comput. infrastructure in the cloud.''\hskip 1em
  plus 0.5em minus 0.4em\relax 2019 {IEEE} Internat. Conf. Fog Comput.
  ({ICFC}), Jun. 2019, pp. 144--152.

\bibitem{Mayer2017-dt}
R.~Mayer, L.~Graser, H.~Gupta, E.~Saurez, and U.~Ramachandran, ``{EmuFog}:
  Extensible and scalable emulation of large-scale fog comput.
  infrastructures.''\hskip 1em plus 0.5em minus 0.4em\relax 2017 {IEEE} Fog
  World Congr. ({FWC}), Oct. 2017, pp. 1--6.

\bibitem{Coutinho2018-wa}
A.~Coutinho, F.~Greve, C.~Prazeres, and J.~Cardoso, ``{Fogbed}: A
  rapid-prototyping emulation environ. for fog computing.''\hskip 1em plus
  0.5em minus 0.4em\relax 2018 {IEEE} Int. Conf. Commun. ({ICC}), May 2018, pp.
  1--7.

\bibitem{Morabito2018-ip}
R.~Morabito, V.~Cozzolino, A.~Y. Ding, N.~Beijar, and J.~Ott, ``Consolidate
  {IoT} edge comput. with lightweight virtualization,'' \emph{IEEE Netw.},
  vol.~32, no.~1, pp. 102--111, 2018.

\bibitem{Govindarajan2014-dm}
N.~Govindarajan, Y.~Simmhan, N.~Jamadagni, and P.~Misra, ``Event process.
  across edge and the cloud for internet of things applications.''\hskip 1em
  plus 0.5em minus 0.4em\relax Proc. 20th Internat. Conf. Manage. Data, Dec.
  2014, pp. 101--104.

\bibitem{Anawar2018-lv}
M.~R. Anawar, S.~Wang, M.~Azam~Zia, A.~K. Jadoon, U.~Akram, and S.~Raza, ``Fog
  computing: An overview of big {IoT} data analytics,'' \emph{Proc. Int. Wirel.
  Commun. Mob. Comput. Conf.}, vol. 2018, pp. 1--22, 2018.

\bibitem{benchmarkingbook}
D.~Bermbach, E.~Wittern, and S.~Tai, \emph{Cloud Service Benchmarking:
  Measuring Qual. of Cloud Services from a Client Perspective}.\hskip 1em plus
  0.5em minus 0.4em\relax Springer, Cham, 2017.

\bibitem{7469991}
W.~Shi and S.~Dustdar, ``The promise of edge computing,'' \emph{Computer},
  vol.~49, no.~5, pp. 78--81, 2016.

\bibitem{diss_bermbach}
D.~Bermbach, ``Benchmarking eventually consistent distrib. storage systems,''
  Ph.D. dissertation, Karlsruhe Institute of Technology, Karlsruhe, Germany,
  Feb. 2014.

\bibitem{kossmann2010evaluation}
D.~Kossmann, T.~Kraska, and S.~Loesing, ``An eval. of alternative architectures
  for transaction process. in the cloud.''\hskip 1em plus 0.5em minus
  0.4em\relax Proc. 2010 ACM SIGMOD Internat. Conf. Manage. Data, Jun. 2010,
  pp. 579--590.

\bibitem{Rausch2020-cm}
T.~Rausch, C.~Lachner, P.~A. Frangoudis, P.~Raith, and S.~Dustdar,
  ``Synthesizing plausible infrastructure configurations for evaluating edge
  comput. systems.''\hskip 1em plus 0.5em minus 0.4em\relax 3rd {USENIX}
  Workshop Hot Topics in Edge Comput. ({HotEdge} 20), Jun. 2020.

\bibitem{Manna2012-cj}
Z.~Manna and A.~Pnueli, \emph{Temporal Verification of Reactive Systems:
  Safety}.\hskip 1em plus 0.5em minus 0.4em\relax Springer Science \& Business
  Media, 2012.

\bibitem{Hasenburg2020-xi}
J.~Hasenburg, F.~Stanek, F.~Tschorsch, and D.~Bermbach, ``Managing latency and
  excess data dissemination in fog-based publish/subscribe systems.''\hskip 1em
  plus 0.5em minus 0.4em\relax Proc. 2nd {IEEE} {International} {Conference}
  {Fog} {Computing} ({ICFC} 2020), Apr. 2020, pp. 9--16.

\bibitem{Mohan2020-cn}
N.~Mohan, L.~Corneo, A.~Zavodovski, S.~Bayhan, W.~Wong, and J.~Kangasharju,
  ``Pruning edge res. with latency shears.''\hskip 1em plus 0.5em minus
  0.4em\relax Proc. 19th {ACM} Workshop Hot Topics in Networks, Nov. 2020, pp.
  182--189.

\bibitem{Skarlat2018-pl}
O.~Skarlat, V.~Karagiannis, T.~Rausch, K.~Bachmann, and S.~Schulte, ``A
  framework for optimization, service placement, and runtime operation in the
  fog.''\hskip 1em plus 0.5em minus 0.4em\relax 2018 {IEEE/ACM} 11th Int. Conf.
  Utility and Cloud Comput. ({UCC}), Dec. 2018, pp. 164--173.

\bibitem{paper_bermbach_benchfoundry}
D.~Bermbach, J.~Kuhlenkamp, A.~Dey, A.~Ramachandran, A.~Fekete, and S.~Tai,
  ``{BenchFoundry}: A benchmarking framework for cloud storage
  services.''\hskip 1em plus 0.5em minus 0.4em\relax {Service-Oriented}
  Computing, Oct. 2017, pp. 314--330.

\bibitem{paper_bermbach_auctions4function_placement}
D.~Bermbach, S.~Maghsudi, J.~Hasenburg, and T.~Pfandzelter, ``Towards
  auction-based function placement in serverless fog platforms.''\hskip 1em
  plus 0.5em minus 0.4em\relax Proc. 2nd {IEEE} {International} {Conference}
  {Fog} {Computing} (ICFC 2020), May 2020, pp. 25--31.

\bibitem{Eisele2017-yx}
S.~Eisele, G.~Pettet, A.~Dubey, and G.~Karsai, ``Towards an architecture for
  evaluating and analyzing decentralized fog applications.''\hskip 1em plus
  0.5em minus 0.4em\relax 2017 {IEEE} Fog World Congr. ({FWC}), Oct. 2017, pp.
  1--6.

\bibitem{Qayyum2018-ml}
T.~Qayyum, A.~W. Malik, M.~A. Khan~Khattak, O.~Khalid, and S.~U. Khan,
  ``{FogNetSim++}: A toolkit for model. and simul. of distrib. fog
  environment,'' \emph{IEEE Access}, vol.~6, pp. 63\,570--63\,583, 2018.

\bibitem{Shekhar2020-yg}
S.~Shekhar, A.~Chhokra, H.~Sun, A.~Gokhale, A.~Dubey, X.~Koutsoukos, and
  G.~Karsai, ``{URMILA}: Dynamically trading-off fog and edge resour. for
  perform. and mobility-aware {IoT} services,'' \emph{Journal Syst.
  Architecture}, vol. 107, no. 101710, 2020.

\bibitem{Heintz2020-lo}
B.~Heintz, A.~Chandra, and R.~K. Sitaraman, ``Optimizing timeliness and cost in
  {Geo-Distributed} streaming analytics,'' \emph{IEEE Trans. Cloud Computing},
  vol.~8, no.~1, pp. 232--245, 2020.

\bibitem{Brogi2017-fk}
A.~Brogi and S.~Forti, ``{QoS-Aware} deployment of {IoT} appl. through the
  fog,'' \emph{IEEE Internet Things Journal}, vol.~4, no.~5, pp. 1185--1192,
  2017.

\bibitem{Tong2016-ke}
L.~Tong, Y.~Li, and W.~Gao, ``A hierarchical edge cloud architecture for mobile
  computing.''\hskip 1em plus 0.5em minus 0.4em\relax {IEEE} {INFOCOM} 2016 -
  The 35th Annu. {IEEE} Internat. Conf. Comput. Communications, Apr. 2016, pp.
  1--9.

\bibitem{Xu2019-rq}
X.~Xu, D.~Li, Z.~Dai, S.~Li, and X.~Chen, ``A heuristic offloading method for
  deep learn. edge services in {5G} networks,'' \emph{IEEE Access}, vol.~7, pp.
  67\,734--67\,744, 2019.

\bibitem{Cardellini2016-gw}
V.~Cardellini, V.~Grassi, F.~Lo~Presti, and M.~Nardelli, ``Optimal operator
  placement for distrib. stream process. applications.''\hskip 1em plus 0.5em
  minus 0.4em\relax Proc. 10th {ACM} Internat. Conf. Distrib. and Event-based
  Systems, Jun. 2016, pp. 69--80.

\bibitem{Oh2020-xj}
K.~Oh, A.~Chandra, and J.~Weissman, ``A netw. cost-aware geo-distributed data
  analytics system.''\hskip 1em plus 0.5em minus 0.4em\relax 2020 20th
  {IEEE/ACM} Internat. Symp. Cluster, Cloud and Internet Comput. ({CCGRID}),
  May 2020, pp. 649--658.

\bibitem{8975987}
S.~Deng, Z.~Xiang, J.~Taheriand, M.~A. Khoshkholghi, J.~Yin, A.~Y. Zomaya, and
  S.~Dustdar, ``Optimal application deployment in resource constrained
  distributed edges,'' \emph{IEEE Transactions on Mobile Computing}, vol.~20,
  no.~5, pp. 1907--1923, 2021.

\bibitem{9132684}
Y.~Ma, W.~Liang, J.~Li, X.~Jia, and S.~Guo, ``Mobility-aware and
  delay-sensitive service provisioning in mobile edge-cloud networks,''
  \emph{IEEE Transactions on Mobile Computing}, 2020.

\bibitem{9305277}
M.~Bagaa, T.~Taleb, J.~B. Bernabe, and A.~Skarmeta, ``Qos and resource-aware
  security orchestration and life cycle management,'' \emph{IEEE Transactions
  on Mobile Computing}, 2020.

\bibitem{9339982}
B.~Nemeth, N.~Molner, J.~Martinperez, C.~J. Bernardos, A.~De~la Oliva, and
  B.~Sonkoly, ``Delay and reliability-constrained vnf placement on mobile and
  volatile 5g infrastructure,'' \emph{IEEE Transactions on Mobile Computing},
  2021.

\bibitem{9358007}
P.~Zhao and G.~Dan, ``Joint resource dimensioning and placement for dependable
  virtualized services in mobile edge clouds,'' \emph{IEEE Transactions on
  Mobile Computing}, 2021.

\bibitem{8955944}
A.~Moubayed, A.~Shami, P.~Heidari, A.~Larabi, and R.~Brunner, ``Edge-enabled
  v2x service placement for intelligent transportation systems,'' \emph{IEEE
  Transactions on Mobile Computing}, vol.~20, no.~4, pp. 1380--1392, 2021.

\bibitem{9373980}
B.~Gao, Z.~Zhou, F.~Liu, F.~Xu, and B.~Li, ``An online framework for joint
  network selection and service placement in mobile edge computing,''
  \emph{IEEE Transactions on Mobile Computing}, 2021.

\bibitem{9361310}
Y.~Li, W.~Dai, X.~Gan, H.~Jin, L.~Fu, H.~Ma, and X.~Wang, ``Cooperative service
  placement and scheduling in edge clouds: A deadline-driven approach,''
  \emph{IEEE Transactions on Mobile Computing}, 2021.

\bibitem{8869806}
Y.~Zhao, X.~Liu, L.~Tu, C.~Tian, and C.~Qiao, ``Dynamic service entity
  placement for latency sensitive applications in transportation systems,''
  \emph{IEEE Transactions on Mobile Computing}, vol.~20, no.~2, pp. 460--472,
  2021.

\bibitem{8960404}
M.~Goudarzi, H.~Wu, M.~Palaniswami, and R.~Buyya, ``An application placement
  technique for concurrent iot applications in edge and fog computing
  environments,'' \emph{IEEE Transactions on Mobile Computing}, vol.~20, no.~4,
  pp. 1298--1311, 2021.

\bibitem{8897679}
H.~Badri, T.~Bahreini, D.~Grosu, and K.~Yang, ``Energy-aware application
  placement in mobile edge computing: A stochastic optimization approach,''
  \emph{IEEE Transactions on Parallel and Distributed Systems}, vol.~31, no.~4,
  pp. 909--922, 2020.

\bibitem{spe.2896}
J.~Paul~Martin, A.~Kandasamy, and K.~Chandrasekaran, ``Crew: Cost and
  reliability aware eagle-whale optimiser for service placement in fog,''
  \emph{Software: Practice and Experience}, 2021.

\bibitem{spe.2951}
Z.~He, K.~Li, K.~Li, and W.~Zhou, ``Server configuration optimization in mobile
  edge computing: A cost-performance tradeoff perspective,'' \emph{Software:
  Practice and Experience}, 2021.

\bibitem{spe.2942}
K.~Cao, T.~Wei, M.~Chen, K.~Li, J.~Weng, and W.~Tan, ``Exploring reliable
  edge-cloud computing for service latency optimization in sustainable
  cyber-physical systems,'' \emph{Software: Practice and Experience}, 2021.

\bibitem{Naas2017-ln}
M.~I. Naas, P.~R. Parvedy, J.~Boukhobza, and L.~Lemarchand, ``{iFogStor}: An
  {IoT} data placement strategy for fog infrastructure.''\hskip 1em plus 0.5em
  minus 0.4em\relax 2017 {IEEE} 1st International Conference on Fog and Edge
  Computing ({ICFEC}), May 2017, pp. 97--104.

\bibitem{Hasenburg2020-yo}
J.~Hasenburg, M.~Grambow, and D.~Bermbach, ``Towards a replication service for
  {Data-Intensive} fog applications.''\hskip 1em plus 0.5em minus 0.4em\relax
  Proceedings of the 35th {ACM} Symposium on Applied Computing, Posters Track
  ({SAC} 2020), 2020.

\bibitem{Hasenburg2019-oe}
------, ``{FBase}: A replication service for {Data-Intensive} fog
  applications,'' TU Berlin \& ECDF, Berlin, Germany, Tech. Rep., 2019.

\bibitem{Khare2019-ef}
S.~Khare, H.~Sun, J.~Gascon-Samson, K.~Zhang, A.~Gokhale, Y.~Barve,
  A.~Bhattacharjee, and X.~Koutsoukos, ``Linearize, predict and place:
  minimizing the makespan for edge-based stream process. of directed acyclic
  graphs.''\hskip 1em plus 0.5em minus 0.4em\relax Proc. 4th {ACM/IEEE} Symp.
  Edge Computing, Nov. 2019, pp. 1--14.

\bibitem{Wobker2018-ud}
C.~W{\"o}bker, A.~Seitz, H.~Mueller, and B.~Bruegge, ``Fogernetes: Deployment
  and manage. of fog comput. applications.''\hskip 1em plus 0.5em minus
  0.4em\relax {NOMS} 2018 - 2018 {IEEE/IFIP} Netw. Operations and Manage.
  Symposium, Apr. 2018, pp. 1--7.

\bibitem{Santos2019-en}
J.~Santos, T.~Wauters, B.~Volckaert, and F.~De~Turck, ``Towards network-aware
  resource provisioning in kubernetes for fog comput. applications.''\hskip 1em
  plus 0.5em minus 0.4em\relax 2019 {IEEE} Conf. Netw. Softwarization
  ({NetSoft}), Jun. 2019, pp. 351--359.

\bibitem{Saurez2016-oo}
E.~Saurez, K.~Hong, D.~Lillethun, U.~Ramachandran, and B.~Ottenw{\"a}lder,
  ``Incremental deployment and migration of geo-distributed situation awareness
  appl. in the fog.''\hskip 1em plus 0.5em minus 0.4em\relax Proc. 10th {ACM}
  Internat. Conf. Distrib. and Event-based Systems, Jun. 2016, pp. 258--269.

\bibitem{Santoro2017-vl}
D.~Santoro, D.~Zozin, D.~Pizzolli, F.~D. Pellegrini, and S.~Cretti, ``Foggy: A
  platform for workload orchestration in a fog comput. environment.''\hskip 1em
  plus 0.5em minus 0.4em\relax 2017 {IEEE} Internat. Conf. Cloud Comput.
  Technol. and Sci. ({CloudCom}), Dec. 2017, pp. 231--234.

\bibitem{Roy2011-uy}
N.~Roy, A.~Dubey, A.~Gokhale, and L.~Dowdy, ``A capacity planning process for
  perform. assurance of component-based distrib. systems.''\hskip 1em plus
  0.5em minus 0.4em\relax Proc. 2nd {ACM/SPEC} Int. Conf. Perform. engineering,
  Sep. 2011, pp. 259--270.

\bibitem{spe.2787}
D.~N. Jha, K.~Alwasel, A.~Alshoshan, X.~Huang, R.~K. Naha, S.~K. Battula,
  S.~Garg, D.~Puthal, P.~James, A.~Zomaya, S.~Dustdar, and R.~Ranjan,
  ``Iotsim-edge: A simulation framework for modeling the behavior of internet
  of things and edge computing environments,'' \emph{Software: Practice and
  Experience}, vol.~50, no.~6, pp. 844--867, 2020.

\bibitem{Brambilla2014-kl}
G.~Brambilla, M.~Picone, S.~Cirani, M.~Amoretti, and F.~Zanichelli, ``A simul.
  platform for large-scale internet of things scenarios in urban
  environments.''\hskip 1em plus 0.5em minus 0.4em\relax Proc. 1st Internat.
  Conf. {IoT} in Urban Space, Oct. 2014, pp. 50--55.

\bibitem{Sotiriadis2014-uz}
S.~Sotiriadis, N.~Bessis, E.~Asimakopoulou, and N.~Mustafee, ``Towards
  simulating the internet of things.''\hskip 1em plus 0.5em minus 0.4em\relax
  2014 28th Internat. Conf. Adv. Inf. Netw. and Appl. Workshops, May 2014, pp.
  444--448.

\bibitem{Zeng2017-lq}
X.~Zeng, S.~K. Garg, P.~Strazdins, P.~P. Jayaraman, D.~Georgakopoulos, and
  R.~Ranjan, ``{IOTSim}: A simulator for analysing {IoT} applications,''
  \emph{Int. J. High Perform. Syst. Archit.}, vol.~72, pp. 93--107, 2017.

\bibitem{Fernandez-Cerero2020-oa}
D.~Fern{\'a}ndez-Cerero, A.~Fern{\'a}ndez-Montes, F.~Javier~Ortega,
  A.~Jak{\'o}bik, and A.~Widlak, ``Sphere: Simulator of edge infrastructures
  for the optim. of perform. and resour. energy consumption,'' \emph{Simulation
  Modelling Pract. and Theory}, vol. 101, no. 1019663, 2020.

\bibitem{Sonmez2018-sq}
C.~Sonmez, A.~Ozgovde, and C.~Ersoy, ``{EdgeCloudSim}: An environ. for perform.
  eval. of edge comput. systems,'' \emph{Trans Emerg. Tel Tech}, vol.~29,
  no.~11, 2018.

\bibitem{Giang2015-ws}
N.~K. Giang, M.~Blackstock, R.~Lea, and V.~C.~M. Leung, ``Developing {IoT}
  appl. in the fog: A distrib. dataflow approach.''\hskip 1em plus 0.5em minus
  0.4em\relax 2015 5th Int. Conf. Internet Things ({IOT}), Oct. 2015, pp.
  155--162.

\bibitem{Banzai2010-lr}
T.~Banzai, H.~Koizumi, R.~Kanbayashi, T.~Imada, T.~Hanawa, and M.~Sato,
  ``{D-Cloud}: Des. of a softw. testing environ. for reliable distrib. syst.
  using cloud comput. technology.''\hskip 1em plus 0.5em minus 0.4em\relax 2010
  10th {IEEE/ACM} Int. Conf. Cluster, Cloud and Grid Computing, May 2010, pp.
  631--636.

\bibitem{De_Oliveira2014-bq}
R.~L.~S. de~Oliveira, C.~M. Schweitzer, A.~A. Shinoda, and L.~Rodrigues~Prete,
  ``Using mininet for emulation and prototyping software-defined
  networks.''\hskip 1em plus 0.5em minus 0.4em\relax 2014 {IEEE} Colombian
  Conf. Commun. and Comput. ({COLCOM}), Jun. 2014, pp. 1--6.

\bibitem{Balasubramanian2014-tj}
D.~Balasubramanian, A.~Dubey, W.~R. Otte, W.~Emfinger, P.~S. Kumar, and
  G.~Karsai, ``A rapid testing framework for a mobile cloud.''\hskip 1em plus
  0.5em minus 0.4em\relax 2014 25nd {IEEE} Int. Symp. Rapid System Prototyping,
  Oct. 2014, pp. 128--134.

\bibitem{Luthra2019-uh}
M.~Luthra and B.~Koldehofe, ``{ProgCEP}: A programming model for complex event
  processing over fog infrastructure.''\hskip 1em plus 0.5em minus 0.4em\relax
  Proceedings of the 2nd International Workshop on Distributed Fog Services
  Design, Dec. 2019, pp. 7--12.

\bibitem{Tuli2019-xs}
S.~MTuli, R.~Mahmud, S.~Tuli, and R.~Buyya, ``{FogBus}: A blockchain-based
  lightweight framework for edge and fog computing,'' \emph{J. Syst. Softw.},
  vol. 154, pp. 22--36, 2019.

\end{thebibliography}

\vfill

\pagebreak
\appendix

\section{Overview of Application Design Options Deployed to Emulated Testbed in Case Study}

\begin{table*}[!h]
  \renewcommand{\arraystretch}{1.3}
  \caption{Overview of the ten most efficient designs as established by our FogExplorer simulation.}
  \label{tab:designs}
  \centering

  \begin{subtable}[h]{\textwidth}
    \resizebox{\textwidth}{!}{
      \begin{tabular}{c|c|c|c|c|c}
        \centering
                                               & \textbf{\emph{A1}}                     & \textbf{\emph{A2}}                     & \textbf{\emph{A3}}                     & \multicolumn{2}{c}{\textbf{\emph{A4}}}                                          \\
        \textbf{\begin{tabular}[c]{@{}c@{}}Application Design\\Option\end{tabular}} & \textbf{\begin{tabular}[c]{@{}c@{}}\emph{Check for Defects}\\Placement\end{tabular}} & \textbf{\begin{tabular}[c]{@{}c@{}}\emph{Adapt Machine}\\Placement\end{tabular}} & \textbf{\begin{tabular}[c]{@{}c@{}}\emph{Predict Pickup}\\Placement\end{tabular}} & \textbf{\begin{tabular}[c]{@{}c@{}}\emph{Aggregate}\\Placement\end{tabular}} & \textbf{\begin{tabular}[c]{@{}c@{}}\emph{Generate Dashboard}\\Placement\end{tabular}} \\ \hline
        1                                      & \emph{FDC}                             & \emph{PKC}                             & \emph{FDC}                             & \emph{WGW}                             & \emph{CLD}                             \\ \hline
        2                                      & \emph{FDC}                             & \emph{FDC}                             & \emph{FDC}                             & \emph{WGW}                             & \emph{CLD}                             \\ \hline
        3                                      & \emph{FDC}                             & \emph{FDC}                             & \emph{ODC}                             & \emph{WGW}                             & \emph{CLD}                             \\ \hline
        4                                      & \emph{FDC}                             & \emph{FDC}                             & \emph{CLD}                             & \emph{WGW}                             & \emph{CLD}                             \\ \hline
        5                                      & \emph{FDC}                             & \emph{PKC}                             & \emph{ODC}                             & \emph{WGW}                             & \emph{FDC}                             \\ \hline
        6                                      & \emph{FDC}                             & \emph{FDC}                             & \emph{ODC}                             & \emph{WGW}                             & \emph{FDC}                             \\ \hline
        7                                      & \emph{FDC}                             & \emph{PKC}                             & \emph{CLD}                             & \emph{WGW}                             & \emph{FDC}                             \\ \hline
        8                                      & \emph{FDC}                             & \emph{FDC}                             & \emph{CLD}                             & \emph{WGW}                             & \emph{FDC}                             \\ \hline
        9                                      & \emph{FDC}                             & \emph{FDC}                             & \emph{FDC}                             & \emph{WGW}                             & \emph{CLD}                             \\ \hline
        10                                     & \emph{FDC}                             & \emph{FDC}                             & \emph{CLD}                             & \emph{WGW}                             & \emph{FDC}                             \\
      \end{tabular}
      \hfill
      \begin{tabular}{r l}
        PKC & Packaging Controller \\
        WGW & Wireless Gateway     \\
        FDC & Factory Data Center  \\
        ODC & Office Data Center   \\
        CLD & Cloud                \\
      \end{tabular}
    }
    \vspace*{5mm}
    \subcaption{Service placement in the different application design options tested on the emulated MockFog testbed.}
  \end{subtable}
  \hfill
  \begin{subtable}[a]{0.45\textwidth}
    \resizebox{\textwidth}{!}{
      \begin{tabular}{c|c|c|c|c}
        \textbf{\begin{tabular}[c]{@{}c@{}}Application Design\\Options\end{tabular}} & \textbf{\begin{tabular}[c]{@{}c@{}c@{}c@{}}\emph{Wireless}\\\emph{Gateway}\\Hardware\\Option\end{tabular}} & \textbf{\begin{tabular}[c]{@{}c@{}c@{}c@{}}\emph{Factory}\\\emph{Data Center}\\Hardware\\Option\end{tabular}} & \textbf{\begin{tabular}[c]{@{}c@{}c@{}c@{}}\emph{Office}\\\emph{Data Center}\\Hardware\\Option\end{tabular}} & \textbf{\begin{tabular}[c]{@{}c@{}c@{}}\emph{Cloud}\\Hardware\\Option\end{tabular}} \\ \hline
        1, 2, 4, 7, 8                          & 1                                      & 2                                       & ---                                     & 1                                       \\ \hline
        3                                      & 1                                      & 2                                       & 1                                       & 1                                       \\ \hline
        5, 6                                   & 1                                      & 2                                       & 1                                       & ---                                     \\ \hline
        9, 10                                  & 2                                      & 2                                       & ---                                     & 1                                       \\
      \end{tabular}
    }
    \vspace*{5mm}
    \subcaption{Infrastructure options in the different application design options tested on the emulated MockFog testbed. Hardware options for the \emph{Camera} and \emph{Production Controller} have been omitted for brevity as no service is deployed on these nodes. }
  \end{subtable}
  \hfill
  \begin{subtable}[a]{0.45\textwidth}
    \resizebox{\textwidth}{!}{
      \begin{tabular}{c|r|r|r|r|r}
                                                & \multicolumn{4}{c|}{\textbf{End-to-End Latency in ms}} &                                                                                                                                                  \\
        \textbf{\begin{tabular}[c]{@{}c@{}}Application Design\\Option\end{tabular}} & \multicolumn{1}{c|}{\textbf{A1}}                       & \multicolumn{1}{c|}{\textbf{A2}} & \multicolumn{1}{c|}{\textbf{A3}} & \multicolumn{1}{c|}{\textbf{A4}} & \textbf{\begin{tabular}[c]{@{}c@{}}Cost in\\\$/month\end{tabular}} \\ \hline
        1                                       & 32.01                                                  & 12                               & 106                              & 266                              & 152.601                                 \\ \hline
        2                                       & 32.01                                                  & 13                               & 106                              & 266                              & 143.101                                 \\ \hline
        3                                       & 32.01                                                  & 13                               & 56                               & 266                              & 173.156                                 \\ \hline
        4                                       & 32.01                                                  & 13                               & 266                              & 266                              & 172.112                                 \\ \hline
        5                                       & 32.01                                                  & 12                               & 56                               & 91                               & 153.055                                 \\ \hline
        6                                       & 32.01                                                  & 13                               & 56                               & 91                               & 143.555                                 \\ \hline
        7                                       & 32.01                                                  & 12                               & 266                              & 91                               & 153.011                                 \\ \hline
        8                                       & 32.01                                                  & 13                               & 266                              & 91                               & 143.511                                 \\ \hline
        9                                       & 32.01                                                  & 13                               & 106                              & 250                              & 178.101                                 \\ \hline
        10                                      & 32.01                                                  & 13                               & 266                              & 75                               & 178.511                                 \\
      \end{tabular}
    }
    \vspace*{5mm}
    \subcaption{Results of the FogExplorer simulation for the ten best application design options tested on the emulated MockFog testbed.}
  \end{subtable}
  \hfill

\end{table*}
\vfill
\pagebreak

\section{Overview of Application Design Options Deployed to the Physical Testbed in Case Study}

\begin{table*}[!h]
  \renewcommand{\arraystretch}{1.3}
  \caption{Overview of the ten application design options selected for deployment on the physical testbed. \emph{W} denotes the most efficient design as determined by our process. \emph{M1-3}, \emph{F1-3}, and \emph{B1-3} denote the three designs that were filtered out by MockFog, FogExplorer, and the application of Best Practices respectively.}
  \label{tab:evaldesigns}
  \centering
  \begin{subtable}[h]{1.0\textwidth}
    \resizebox{\textwidth}{!}{
      \begin{tabular}{c|c|c|c|c|c}
        \centering
                                           & \textbf{\emph{A1}}                 & \textbf{\emph{A2}}                 & \textbf{\emph{A3}}                 & \multicolumn{2}{c}{\textbf{\emph{A4}}}                                      \\
        \textbf{\begin{tabular}[c]{@{}c@{}}Application Design\\Option\end{tabular}} & \textbf{\begin{tabular}[c]{@{}c@{}}\emph{Check for Defects}\\Placement\end{tabular}} & \textbf{\begin{tabular}[c]{@{}c@{}}\emph{Adapt Machine}\\Placement\end{tabular}} & \textbf{\begin{tabular}[c]{@{}c@{}}\emph{Predict Pickup}\\Placement\end{tabular}} & \textbf{\begin{tabular}[c]{@{}c@{}}\emph{Aggregate}\\Placement\end{tabular}}     & \textbf{\begin{tabular}[c]{@{}c@{}}\emph{Generate Dashboard}\\Placement\end{tabular}} \\ \hline
        W                                  & \emph{FDC}                         & \emph{PKC}                         & \emph{ODC}                         & \emph{WGW}                             & \emph{FDC}                         \\ \hline
        M1                                 & \emph{FDC}                         & \emph{PKC}                         & \emph{FDC}                         & \emph{WGW}                             & \emph{CLD}                         \\ \hline
        M2                                 & \emph{FDC}                         & \emph{FDC}                         & \emph{CLD}                         & \emph{WGW}                             & \emph{CLD}                         \\ \hline
        M3                                 & \emph{FDC}                         & \emph{PKC}                         & \emph{CLD}                         & \emph{WGW}                             & \emph{FDC}                         \\ \hline
        F1                                 & \emph{FDC}                         & \emph{FDC}                         & \emph{ODC}                         & \emph{WGW}                             & \emph{CLD}                         \\ \hline
        F2                                 & \emph{FDC}                         & \emph{PKC}                         & \emph{CLD}                         & \emph{WGW}                             & \emph{CLD}                         \\ \hline
        F3                                 & \emph{WGW}                         & \emph{PKC}                         & \emph{FDC}                         & \emph{WGW}                             & \emph{ODC}                         \\ \hline
        B1                                 & \emph{PKC}                         & \emph{CLD}                         & \emph{PRC}                         & \emph{ODC}                             & \emph{FDC}                         \\ \hline
        B2                                 & \emph{CLD}                         & \emph{ODC}                         & \emph{WGW}                         & \emph{WGW}                             & \emph{PKC}                         \\ \hline
        B3                                 & \emph{FDC}                         & \emph{CLD}                         & \emph{FDC}                         & \emph{PKC}                             & \emph{PKC}                         \\
      \end{tabular}
      \begin{tabular}{r l}
        PRC & Production Controller \\
        PKC & Packaging Controller  \\
        WGW & Wireless Gateway      \\
        FDC & Factory Data Center   \\
        ODC & Office Data Center    \\
        CLD & Cloud                 \\
      \end{tabular}
    }
    \vspace*{5mm}
    \subcaption{Service placement in the different application design options tested on the physical testbed.}
  \end{subtable}

  \begin{subtable}[h]{0.45\textwidth}
    \resizebox{\textwidth}{!}{
      \begin{tabular}{c|c|c|c|c}
        \textbf{\begin{tabular}[c]{@{}c@{}}Application Design\\Options\end{tabular}} & \textbf{\begin{tabular}[c]{@{}c@{}c@{}c@{}}\emph{Wireless}\\\emph{Gateway}\\Hardware\\Option\end{tabular}} & \textbf{\begin{tabular}[c]{@{}c@{}c@{}c@{}}\emph{Factory}\\\emph{Data Center}\\Hardware\\Option\end{tabular}} & \textbf{\begin{tabular}[c]{@{}c@{}c@{}c@{}}\emph{Office}\\\emph{Data Center}\\Hardware\\Option\end{tabular}} & \textbf{\begin{tabular}[c]{@{}c@{}c@{}}\emph{Cloud}\\Hardware\\Option\end{tabular}} \\ \hline
        W                                  & 1                                  & 2                                   & 1                                   & ---                                 \\ \hline
        M1, M2, M3                         & 1                                  & 2                                   & ---                                 & 1                                   \\ \hline
        F1                                 & 1                                  & 2                                   & 2                                   & 3                                   \\ \hline
        F2                                 & 1                                  & 2                                   & 3                                   & 3                                   \\ \hline
        F3                                 & 1                                  & 2                                   & 3                                   & ---                                 \\ \hline
        B1                                 & ---                                & 2                                   & 1                                   & 2                                   \\ \hline
        B2                                 & 1                                  & ---                                 & 1                                   & 3                                   \\ \hline
        B3                                 & ---                                & 2                                   & ---                                 & 1                                   \\
      \end{tabular}
    }
    \vspace*{5mm}
    \subcaption{Infrastructure options in the different application design options tested on the physical testbed. Hardware options for the \emph{Camera} and \emph{Production Controller} have been omitted for brevity as no service is deployed on these nodes.}
  \end{subtable}
\end{table*}

\end{document}